\documentclass{amsart}
\usepackage{amssymb,stmaryrd}
\usepackage{amsfonts}
\usepackage{amstext}
\usepackage{algorithmic}
\usepackage{algorithm}
\usepackage{graphicx}
\usepackage{epstopdf}
\usepackage[all]{xy}
\usepackage{MnSymbol}
\usepackage{tikz}
\usepackage{braket}

\numberwithin{equation}{section}
\newtheorem{theorem}{Theorem}[section]
\newtheorem{lemma}[theorem]{Lemma}
\newtheorem{proposition}[theorem]{Proposition}
\newtheorem{corollary}[theorem]{Corollary}

\theoremstyle{definition}

\theoremstyle{remark}

\newcommand{\R}{{\mathbb{R}}}

\newcommand{\C}{{\mathbb{C}}}
\newcommand{\Z}{{\mathbb{Z}}}

\newcommand{\F}{{\mathbb{F}}}

\newcommand{\<}{{\langle}}
\renewcommand{\(}{{(}}
\renewcommand{\)}{{)}}
\renewcommand{\>}{{\rangle}}

\newcommand{\CC}{{\mathcal{C}}}

\newcommand{\cg}{{\mathfrak{g}}}
\newcommand{\Gn}{{G}}
\newcommand{\Gp}{{\mathbb{G}}}

\newcommand{\wedgeq}{{\wedge\kern-5pt\cdot}}

\newcommand{\tens}{\otimes}

\newcommand{\id}{{\rm id}}

\newcommand{\extd}{{\rm d}}
\newcommand{\del}{{\partial}}

\newcommand{\twoForm}[2]{e^{#1} \otimes e^{#2}}
\newcommand{\sigFun}[2]{ \sigma(\twoForm{#1}{#2}) }

\newcommand{\cp}{R}

\begin{document}

\title{Quantum gravity on polygons and $\R\times \Z_n$ FLRW model}
\keywords{noncommutative geometry, quantum groups, quantum gravity, expanding universe, discrete model, Friedmann Robertson Walker, cosmology, Hawking effect}

\subjclass[2000]{Primary 81R50, 58B32, 83C57}

\author{J. N. Argota-Quiroz and S. Majid}
\address{Queen Mary, University of London\\
School of Mathematics, Mile End Rd, London E1 4NS, UK}

\email{j.n.argotaquiroz@qmul.ac.uk, s.majid@qmul.ac.uk}

\thanks{Ver 1.7. The first author was partially supported by CONACyT (M\'exico)}
\date{}

\begin{abstract}We fully solve the quantum geometry of $\Z_n$ as a polygon graph with arbitrary metric square-lengths on the edges, finding a  $*$-preserving quantum Levi-Civita connection which is unique for $n\ne 4$. As a first application, we numerically compute correlation functions for Euclideanised quantum gravity on $\Z_n$ for small $n$. We then study an FLRW model on $\R\times\Z_n$, finding the same expansion rate as for the classical flat FLRW model in 1+2 dimensions.   We also look at particle creation on $\R\times \Z_n$ and find an additional $m=0$ adiabatic no particle creation expansion as well as the particle creation spectrum for a smoothed step expansion.   \end{abstract}
\maketitle 

\section{Introduction}

This paper is the third in a sequence\cite{Ma:haw,Ma:sq} in which we apply the recently developed formalism of `quantum Riemannian geometry'\cite{BegMa:book,Ma:gra, BegMa:gra,MaTao,MaPac} to quantum gravity on discrete spacetimes, as a new approach compared to previous lattice and other discrete schemes \cite{Loll,Ash,Dow,Hale}. It is also very different from loop quantum cosmology such as \cite{Ash} but has in common the idea of hugely simplifying the gravitational degrees of freedom to the point of something calculable. In the first of the two papers, we solved quantum gravity on a square\cite{Ma:sq} but with differential structure given by the group $\Z_2\times\Z_2$, which is different from the case of a polygon based on the group $\Z_n$ that we treat now, even when $n=4$. The metric data in our approach consists of generic `square length' data assigned to each edge of the polygon, while the group structure affects the choice of 2-forms and higher forms, but also provides a natural basis for the 1-forms in the same way that one implicitly fixes a differential structure classically by use of  translation-invariant differentials $\extd x^\mu$.   In \cite{Ma:haw}, we similarly set up but could not solve quantum gravity on the lattice line $\Z$, but did manage computations for cosmological particle creation on a universe with only a lattice time direction. The connection and Einstein-Hilbert action now on $\Z_n$ will turn out, which is the main result of Section~\ref{secpoly}, to be a modulo $n$ version of the ones for $\Z$. This is not surprising but required significant proof and in fact turns out to be not the only possibility for $n=4$. The main difference, however, is that now, for finite $n$ depending only on computing power, Euclidean quantum gravity {\em is} fully computable, as we demonstrate in Section~\ref{secqg} for $n\le 6$. We switched here to a Euclidean interpretation since a periodic time direction makes no sense physically, but we note that Euclidean quantum gravity is still of interest for classical compact Riemannian manifolds without boundary, see \cite{HawGib}. Also note that the overall  normalisation of the metric does not enter into the connection $\nabla$ nor into the quantum gravity action, but from a positivity point of view, since our basis of invariant one-forms obey $e^+{}^*=-e^-$, the inverse metric in Section~\ref{secpoly} is actually negative definite in the sense $(e^+,e^+{}^*)=-{1\over a}<0$ for the associated hermitian metric evaluated on the diagonal and with `square-length' function $a>0$. Section~\ref{seccirc} looks carefully at the $n\to \infty$ limit of the polygon and identifies it as a central extension in the sense of \cite{Ma:alm,Ma:rec} of the classical calculus on a circle, with an extra `normal' direction $\Theta_0$. 

Our second main result is then a detailed study of the FLRW model on $\R\times\Z_n$ with $\R$ classical, including  cosmological particle creation following the approach of Parker \cite{Par69,Par12,ParNav,ParTom}. This is an important part of quantum theory on curved spacetime and relates also to Bekenstein-Hawking radiation, see  \cite{MukWin,Birrel}.  We first find in Section~\ref{seccross} that quantum metrics on $\R\times\Z_n$ are forced to have the block form $g=\mu\extd t\tens\extd t+ h_{ab}e^a\tens e^b$ and, moreover, $h_{ab}$ has to have a specific form where the time dependence enters uniformly in the spatial metric.  Section~\ref{secFRW} focusses on the FLRW cosmology case of a uniform metric on $\Z_n$  expanded by a time-dependent factor $\cp(t)$, so 
\begin{equation} g=-\extd t\tens \extd t- \cp^2(t)(e^+\tens e^-+e^-\tens e^+).\end{equation}
The negative sign  in the second term is required by our positivity remark above and we then find that the Friedmann equations for $\cp(t)$ in our discrete case actually come out the same as for the usual flat FLRW model in {\em two} spatial dimensions, which is in line with our cotangent bundle on $\Z_n$ being necessarily 2-dimensional, not 1-dimensional.   Section~\ref{secqftR} provides some elementary checks for QFT in the constant $\cp$ case, then Section~\ref{sechawR} covers the cosmological particle creation for varying $\cp(t)$.  We first consider the classical geometry case of $\R\times S^1$, which sets up the formalism and for which we did not find a suitable treatment elsewhere, then the modifications for $\R\times \Z_n$. Of interest are the adiabatic no particle creation possibilities for $\cp(t)$ aside from the obvious constant $\cp$ case; for $\R\times S^1$ there is a further possibility with $m\to \infty$, but for  $\R\times\Z_n$  we find a second further possibility with $m\to 0$. The particle creation calculation itself is done only for  `in' and `out' regimes of constant $\cp$, with results a little different in the $\Z_n$ case due to the periodic nature of the spatial momentum compared to the $S^1$ case, see Figure~\ref{fig:pol-circle}. 

Since this is the third in a sequence of papers, we keep the remaining general remarks, as well as the recap of the formalism in the preliminaries Section~\ref{secpre}, to a minimum. Suffice it to say that our particular approach to discrete quantum gravity has its roots in {\em quantum spacetime} or the idea that space and time coordinates are noncommutative or `quantum'. This was speculated on since the early days of quantum theory but has also emerged by now as a better-than-classical effective theory that includes some quantum gravity effects. It was first discussed in modern times in \cite{Ma:pla} in the context of non-commutativity of phase space  and quantum Born reciprocity or observable-state duality, where it led to the bicrossproduct class of quantum groups (rather different from the other main class, the $q$-deformation ones, arising from  integrable systems). An important model here was the bicrossproduct model Minkowski spacetime $[x_i,t]=\imath\lambda_p x_i$ in \cite{MaRue}, with quantum group Poincar\'e symmetry having a bicrossproduct form (as well as a construction\cite{Luk} by contraction from $U_q(so_{3,2})$). Other early works were \cite{Sny}, which did not itself propose a closed spacetime algebra, its adaptation \cite{DFR} with classical (not quantum) symmetry and the proposal \cite{Hoo} of the angular momentum algebra as a quantum spacetime. We refer to \cite{Ma:sq} for more details and literature.  What is important is that, as argued back in  \cite{Ma:pla},  a true theory of quantum gravity effects on spacetime also needed  models where the spacetime (and indeed the entire position-momentum space) was both curved and quantum, and for this one needed an actual formalism for that. 

What emerged, somewhat different in character from  `noncommutative geometry \`a la Connes'\cite{Con} coming out of cyclic cohomology,  K-theory and `spectral triples' as abstract Dirac operators,  was a more  constructive `quantum groups approach' motivated by quantum groups and their homogeneous spaces as  examples but ultimately working for any algebra $A$ equipped with differential structure, over any field. The starting point here is to specify the differential structure as a bimodule $\Omega^1$ of `1-forms' (this means we can multiply them from either side by elements of $A$) equipped with an exterior derivative $\extd: A\to \Omega^1$ obeying the Leibniz rule. We then define a metric as an invertible element  $g\in\Omega^1\tens_A\Omega^1$ with some kind of symmetry condition and a quantum Levi-Civita connection (QLC) in these terms is a bimodule connection\cite{DVM,Mou} $\nabla:\Omega^1\to \Omega^1\tens_A\Omega^1$ which is metric compatible and torsion free. For each quantum Riemannian geometry, one can compute a Laplacian $\Delta=(\ ,\ )\nabla\extd: A\to A$ and, with a little more `lifting' data, a Ricci tensor in $\Omega^1\tens_A\Omega^1$ and a Ricci scalar $S\in A$, see Section~\ref{secpre} and \cite[Chap.~8.1]{BegMa:book} for more details and references. Along with this new formalism has come a new generation of examples. Notably, the above bicrossproduct model spacetime in 1+1 dimensions turned out  \cite{BegMa:gra,MaTao} from this point of view to  admit two main classes of translation invariant 2D differential structures and each of these to admit a 1-parameter moduli of curved quantum Riemannian geometries. 

Finally, which is the critical thing for discrete quantum gravity, this emergent quantum Riemannian geometry, since it works for any algebra, can just as well be applied to functions on a discrete set. Here the algebra is commutative but it turns out that differentials $\Omega^1$ on this algebra are the same thing as graphs with the given set as vertices, and cannot commute with functions for consistency of the Leibniz rule. So this is a very different regime  from deformation-type quantum spacetimes, but the thinking is the same; a  better model of spacetime that includes some quantum gravity effects, reflected now in discrete positions and  noncommutative differentials. More details are in Section~\ref{secpre} and \cite{Ma:haw,Ma:sq,BegMa:book,Ma:gra}. The algebra $A$ can also perfectly well be finite-dimensional, so in our case this immediately includes  calculable `baby quantum gravity' models where spacetime is a small graph. What is critical for this to have any validity is that the theory is not ad-hoc to the discrete case but simply one extreme end of a single functorial framework that include continuum geometries and their deformations at the other end. 

Section~\ref{secRem} concludes with some directions for further work. We work in units with $\hbar=c=1$. 

\section{Preliminaries} \label{secpre}

As mentioned, it is important that we have a single formalism that includes both classical and discrete cases as well as the mixed case of $\R\times\Z_n$ needed in the paper. For each layer of Riemannian geometry, we briefly recall the general set up over a unital algebra $A$ as in \cite{BegMa:book}, for orientation purposes, then give details for the discrete graph case where $A$ is functions on a discrete set, which was also the setting of \cite{Ma:haw,Ma:sq}. 

\subsection{Differentials and metrics}  As explained in the introduction, the first step is a graded exterior algebra $(\Omega,\extd)$ where $\Omega^0=A$ is the algebra of `functions' and $\extd$ increases the differential form degree by 1, obeys a graded-Leibniz rule and $\extd^2=0$. We also require $\Omega$ to be generated by $A,\extd A$. If one fixes $\Omega^1$ first then there is a unique `maximal prolongation' $\Omega$ of which one can chose a quotient if one wants. In our case, we will be interested in the commutative algebra $A=C(X)$ of complex functions on a discrete set $X$ with pointwise product. Then choosing $\Omega^1$ is equivalent to assigning arrows to make a graph with vertex set $X$. Denoting a vector space  basis of $\Omega^1$ by $\{\omega_{x\to y}\}$ labelled by arrows $x\to y$, the bimodule products and exterior derivative are
\begin{equation}  f.\omega_{x\to y}=f(x)\omega_{x\to y},\quad \omega_{x\to y}.f=f(y)\omega_{x\to y},\quad \extd f=\sum_{x\to y}(f(y)-f(x))\omega_{x\to y}.\end{equation}
We will be interested in the case where the graph is bidirected i.e., for every arrow $x\to y$ there is an arrow $y\to x$. In other words, the data is just a usual undirected graph which we understand as arrows both ways in the above formulae. A metric as a tensor $g\in \Omega^1\tens_A\Omega^1$  then has the form
\begin{equation} g=\sum_{x\to y}g_{x\to y}\omega_{x\to y}\tens\omega_{y\to x}\in \Omega^1\tens_{C(X)}\Omega^1\end{equation}
for nonzero weights $g_{x\to y}$ for every edge, as is dictated by being central \cite[Chap.~1.4]{BegMa:book} \cite{Ma:gra}. Here a term $\omega_{x\to y}\tens\omega_{y'\to x'}$ needs $y'=y$ to be nonzero since we could left-multiply the second factor  by $\delta_{y'}$ which does not change it, or move $\delta_{y'}$ over to the first factor where it acts from the right by $\delta_{y',y}$. And it needs $x'=x$ to be central so that right multiplication by $\delta_{x'}$ on the second factor, which does not change it, can coincide with left-multiplication on the first factor, which gives $\delta_{x',x}$. That centrality is needed for bimodule invertibility was a key result in  \cite{BegMa:gra}. Canonically, a metric is `quantum symmetric'  if $\wedge(g)=0$ for the wedge product of $\Omega$. Specific to graphs, we also have a slightly different notion that $g$ is {\em edge-symmetric} if $g_{x\to y}=g_{y\to x}$ for all $x\to y$, i.e., does not depend on the direction of travel. As in \cite{Ma:haw} for the line graph, we will see that this variant also works better when we apply it to the polygon. 
 
Next, it is useful to endow $X$ with a group structure and look for $\Omega^1$ which is left and right translation invariant. These will be the Cayley graph for an Ad-stable set of generators $\CC\subseteq \Gp \setminus\{e\}$ (where $e$ is the  identity of a group $\Gp$), with arrows of the form $x\to xa$ for $a\in \CC$. In this case, one has a basis of invariant 1-forms $ e^a=\sum_{x\to x a} \omega_{x\to xa}$ with $\Omega^1=A.\{e^a\}$, bimodule relations and derivative
\begin{equation} e^a f= R_a(f)e^a,\quad \extd f=\sum_a (\del_a f)e^a,\quad \del_a=R_a-\id,\quad R_a(f)(x)=f(xa)\end{equation}
defined by the right translation operators $R_a$ as stated. These formulae now makes sense even when $X$ is infinite as long as $\CC$ is finite. Moreover  $\Omega$ is canonically generated  by functions and basic 1-forms with the above as well as  certain  `braided-anticommutation relations' between the $\{e^a\}$. In the case of an Abelian group (which is all we will need), this is just the usual  Grassmann algebra on the $e^a$, i.e., they anticommute and we also have $\extd e_a=0$ in this case.  

\subsection{Connections}  A connection in quantum Riemannian geometry is a map $\nabla:\Omega^1\to \Omega^1\tens_A\Omega^1$. Given a quantum vector field in the form of a right module map $X:\Omega^1\to A$, we can evaluate this against the first output to obtain a covariant derivative $\nabla_X:\Omega^1\to \Omega^1$, but the connection itself is defined independently of any vector field. It is required to obey two Leibniz rules as follows. From the left, we ask for
\begin{equation} \nabla(a\omega)=\extd a\tens\omega+ a\nabla\omega\end{equation}
for all $a\in A,\omega\in \Omega^1$. From the right, we similarly ask\cite{DVM,Mou} for
\begin{equation} \nabla(\omega a)=(\nabla\omega)a+\sigma(\omega\tens\extd a);\quad \sigma:\Omega^1\tens_A\Omega^1\to \Omega^1\tens_A\Omega^1\end{equation}
for some `bimodule map' $\sigma$ (i.e. commuting with the action of $A$ from either side, so `tensorial' in a strong sense.) 

In the case we need of a Cayley graph calculus on a group, we see that $\nabla$ just needs to be specified on the $e^a$ provided this is consistent with its  extension to $\Omega^1$ by the two Leibniz rules. We write
\begin{equation} \nabla e^a=-\Gamma^a{}_{bc}e^b\tens e^c,\quad \sigma(e^a\tens e^b)=\sigma^{ab}{}_{mn}e^m\tens e^n\end{equation}
for coefficients in $A$ with a certain compatibility between these tensors for a bimodule connection. In general, torsion-free amounts to $\wedge\nabla-\extd=0$ as maps from $\Omega^1\to \Omega^2$ and needs in the case of an abelian group the additional relations
\begin{equation} \Gamma^a{}_{bc}=\Gamma^a{}_{cb},\quad \sigma^{ab}{}_{mn}e^m\wedge e^n+ e^a\wedge e^b=0.\end{equation}

Next, any bimodule connection extends canonically to a connection on tensor products. This implies a meaning to $\nabla g=0$, namely if $g=g^1\tens g^2$, say, then this is 
\begin{equation} \nabla g^1\tens g^2+(\sigma(g^1\tens(\ ))\tens\id)\nabla g^2=0.\end{equation}
In the discrete Cayley graph setting, we write $g=h_{ab}e^a\tens e^b$, where centrality needs \begin{equation} h_{ab}=\delta_{a^{-1},b}h_a\end{equation}
 for some functions $h_a$. In these terms, (i) edge symmetry and, in the case of the Grassmann algebra, quantum symmetry (ii) appear as
\begin{equation} (i)\quad h_a=R_a(h_{a^{-1}}),\quad (ii)\quad  h_a=h_{a^{-1}}.\end{equation}

\subsection{*-structures, inner calculi and structure constants} For physics, there should also be a $*$-involution on $A$, which in our examples is just pointwise complex conjugation, and everything should be unitary or `real' in the sense of $*$-preserving. We require this to extend to $\Omega$ with an extra minus signs for swapping two odd elements and to commute with $\extd$. For the metric and connection, `reality' means
\begin{equation}\label{nablastar} g^\dagger=g,\quad \nabla \circ *= \sigma\circ\dagger\circ\nabla, \end{equation}
which also implies $\dagger\circ\sigma=\sigma^{-1}\circ\dagger$ with $(\omega\tens\eta)^\dagger=\eta^*\tens\omega^*$ for $\omega,\eta\in \Omega^1$. In the Cayley graph case, $e^a{}^*=-e^{a^{-1}}$ is the natural choice. Then reality of the metric is  $\overline{h_{ab}}=h_{b^{-1},a^{-1}}$, which means the metric functions $h_a$ are real valued. For $\Gamma$, the formula depends on $\sigma$ and is more complicated. 

Finally, when the calculus is inner in the sense of a 1-form $\Theta$ which generates $\extd$ by graded-commutator $\extd=[\Theta,\ \}$, it is shown in  \cite{Ma:gra} that 
\begin{equation} \nabla \omega = \Theta\otimes \omega  + (\alpha-\sigma_\Theta)(\omega)\end{equation}
for some bimodule map $\alpha:\Omega^1\to \Omega^1\tens_A\Omega^1$ and some bimodule map $\sigma$, with $\sigma_\Theta=\sigma((\ )\tens\Theta)$. To be torsion free, we require the condition on $\sigma$ as above and $\wedge\alpha=0$. To be metric compatible, we need
\begin{equation} \Theta\tens g+(\alpha\tens\id)g+(\sigma\tens\id)(\id\tens (\alpha-\sigma_\Theta))g=0.\end{equation}
To be `real', we need the condition on $\sigma$ above and $\alpha\circ *=\sigma\circ\dagger\circ\alpha$.  A Cayley graph calculus is inner with $\Theta=\sum_a e^a$. In this case, to be bimodule maps, we need $\sigma^{ab}{}_{mn}=0$ unless $ab=mn$ in the group and $\alpha(e^a)=\alpha^a{}_{mn}e^m\tens e^n$ needs $\alpha^a{}_{mn}=0$ unless $a=mn$ in the group, see \cite[Chap. 8.2.2]{BegMa:book}\cite{Ma:gra}. The indices here range over elements of the generating set $\CC$ of the calculus and are {\em not} being multiplied in the 4-index and 3-index tensors $\sigma^{ab}{}_{mn},\alpha^a{}_{mn}$. We will need this a little more explicitly than currently in the literature. 

\begin{lemma}\label{cayleystr} Let $\Omega(\Gp )$ be a Cayley graph calculus and cf.   \cite{BegMa:book,Ma:gra}, write a bimodule connection on $\Omega^1$ in the form
\[ \sigma^{ab}{}_{mn}=\delta^a{}_n\delta^b{}_m+\delta^b{}_{a^{-1}mn}\tau^{a}{}_{mn},\quad \Gamma^a{}_{bc}=\tau^a{}_{bc}-\delta^a{}_{bc}\alpha_{bc}\]
for coefficient functions $\tau^a{}_{bc}=0$ unless $a^{-1}bc\in\CC$ and $\alpha_{bc}=0$ unless $bc\in\CC$.

(1) For $\Gp $ abelian, the condition for torsion freeness is $\tau^a{}_{bc},\alpha_{bc}$ symmetric in $b,c$. 

(2) The conditions for `reality' of the connection (to be $*$-preserving) are
\[  \alpha_{bc}+R_{bc}(\overline{\alpha_{c^{-1}b^{-1}}}) +\sum_n R_{nbcn^{-1}}(\overline {\alpha_{c^{-1}b^{-1}n^{-1},n}})\tau^{n^{-1}}{}_{bc}=0,\] \[\tau^{a^{-1}}{}_{cd}+R_{cd}(\overline{\tau^a{}_{c^{-1}d^{-1}}})+\sum_n R_{cd}(\overline{\tau^a{}_{c^{-1}d^{-1}n,n^{-1}}})\tau^n{}_{cd}=0 \]
for all $a,b,c,d$.

(3) The conditions for metric compatibility with an edge-symmetric metric are
\[  h_{mn}\alpha_{mn}+R_n(h_{n^{-1}}\alpha_{m,n^{-1}m^{-1}})- \sum_a R_{a^{-1}}(h_a\alpha_{amn,n^{-1}m^{-1}}) -R_n(h_{n^{-1}}\tau^{n^{-1}}{}_{m,n^{-1}m^{-1}})=0, \]
\[ \delta^p{}_{n^{-1}}\del_m h_n=h_{p^{-1}}\tau^{p^{-1}}{}_{mn}-\sum_a R_{a^{-1}}(h_a\tau^a{}_{amn,p})\tau^{a^{-1}}{}_{mn}\] 
for all $m,n,p$. 
\end{lemma}
\proof (1) The first formula displayed is basically in \cite{Ma:gra} (or see \cite[Chap. 8.2.2]{BegMa:book}) in the inner case with $\Theta=\sum_a e^a$, merely put in terms of the components of $\Gamma$ and after subtracting off the flip map from $\sigma$ and imposing the bimodule properties of the maps $\alpha,\sigma$ (hence written in terms of $\tau$).  It is easy to see that $\wedge\alpha=0$ and $\wedge(\id+\sigma)=0$ for the Grassmann algebra case reduce to symmetry in the lower indices (this technique is used in \cite{BegMa:book} but is in any case straightforward). Note that $e\notin\CC$ so $\Gamma^a{}_{bc}$ has value $-\alpha_{bc}:=-\alpha_{b,c}$ when $a=bc$ and $\tau^a{}_{bc}:=\tau^a{}_{b,c}$ otherwise. We omit the commas when there are only two elements not being multiplied. 

(2) The condition for $\alpha$ is immediate from $\sigma\circ\dagger\circ\alpha=\alpha\circ *$ evaluated on $e^a$ with $e^a{}^*=-e^{a^{-1}}$. The condition $\sigma\circ\dagger\circ\sigma=\dagger$  is easily seen (as in the proof of \cite[Lemma~8.17]{BegMa:book} for $\alpha=0$) to be
\begin{equation} \sum_{m,n}R_{n^{-1}m^{-1}}(\overline{\sigma^{ab}{}_{mn}})\sigma^{n^{-1}m^{-1}}{}_{cd}=\delta^{b^{-1}}{}_c\delta^{a^{-1}}{}_d,\end{equation}
which we now evaluate for the stated form of $\sigma$. 

(3)  Metric compatibility is 
\begin{equation} \nabla(h_{ab}e^a)\tens e^b-\sigma(h_{ab}e^a\tens \Gamma^b{}_{cd}e^c)\tens e^d=0,\end{equation}
which expands out using the Leibniz rules and the form of the metric to
\begin{equation} \delta_{p,n^{-1}}\del_m h_n-h_{p^{-1}}\Gamma^{p^{-1}}{}_{mn}-h_a R_a(\Gamma^{a^{-1}}{}_{bp})\sigma^{ab}{}_{mn}=0\end{equation}
In the edge-symmetric case, this becomes 
\begin{equation} \delta_{p,n^{-1}}\del_m h_n-h_{p^{-1}}\Gamma^{p^{-1}}{}_{mn}- R_a(h_{a^{-1}}\Gamma^{a^{-1}}{}_{bp})\sigma^{ab}{}_{mn}=0.\end{equation}
We now insert the form of $\Gamma,\sigma$ to obtain the condition stated in the mutually exclusive cases $p=n^{-1}m^{-1}$ and  $p\ne n^{-1}m^{-1}$ (where the terms shown do not contribute when $p=n^{-1}m^{-1}$ due to the conditions on $\tau$ and $e\notin\CC$, so we do not need to write that this $p$ is excluded). \endproof

We will apply this to $\Gp =\Z_n$ and $\CC=\{\pm1\}$, denoting the corresponding basis indices for brevity as $\pm$.  The Cayley graph is then the polygon with arrows in both directions.

\subsection{Curvature}\label{seccurv} Given a left connection $\nabla$ on an algebra with differential calculus (it does not even need to be a bimodule one), we have Riemann curvature
\begin{equation} R_\nabla:\Omega^1\to \Omega^2\tens_A\Omega^1,\quad R_\nabla=(\extd\tens\id-\id\wedge\nabla)\nabla.\end{equation}
For example, in the inner case of a connection defined by maps $\sigma,\alpha$ as above, this is 
\begin{equation} R_{\nabla}\omega = \Theta \wedge \Theta \otimes \omega - (\wedge \otimes  \id)(\id\otimes(\alpha-\sigma_\Theta))(\alpha - \sigma_\Theta)\omega\end{equation}
for all $\omega\in\Omega^1$.

Next, given a bimodule `lift' map $i:\Omega^2\to \Omega^1\tens_A\Omega^1$ such that $\wedge\circ i=\id$, we define Ricci and the Ricci scalar $S$ relative to it as
\begin{equation} {\rm Ricci} = ((\ ,\ ) \otimes \id)(\id\otimes i \otimes \id)(\id\otimes R_\nabla)g,\quad 
	S = (\ ,\ ){\rm Ricci}. \end{equation}
This is a `working definition' rather than part of a fully developed theory (for which an understanding of conservation laws and the stress-energy tensor would be needed). In the Cayley graph case of Lemma~\ref{cayleystr}, there is a canonical $\Omega$ and with it a canonical $i$ in \cite[Lem.~8.18]{BegMa:book}, which for an abelian group is just 
\begin{equation} i(e^a\wedge e^b)={1\over 2}(e^a\tens e^b- e^b\tens e^a)\end{equation}
on the Grassmann algebra generators (extended as a bimodule map). Thus, once we have found a QLC for our quantum metric, the route to the scalar curvature needed for the Einstein-Hilbert action is canonical at least for Abelian groups such as $\Z_n$.

\section{Quantization of  $\Z_n$}

Here we consider the general theory above for the case of an $n$-gon for $n\ge 3$. A metric is a free assignment of a `square-length' to each edge and Section~\ref{secpoly} solves the quantum Riemannian geometry to find the quantum Levi-Civita connection. Section~\ref{secqg} then constructs  Euclidean quantum gravity on the polygon. 

\subsection{Quantum Riemannian geometry on  $\Z_n$}\label{secpoly} Just as it is useful in classical geometry to use local coordinates where the differential structure is the standard one for $\R^n$, it is similarly useful to regard the $n$-gon as the group $\Gp =\Z_n$ for its differential structure as explained in Section~\ref{secpre}. Here the calculus $\Omega^1 (\Z_n)$ with generators $\CC = \{1,-1\}$ has corresponding left-invariant basis $e^+,e^-$ given by
\begin{equation}e^+ = \sum_{i=0}^{n-1} \omega_{i \rightarrow i+1}; \quad e^- = \sum_{i=0}^{n-1} \omega_{i \rightarrow i-1},
\end{equation}
where $i\in\Z_n$ runs over the vertices. The $n=2$ case is different and was already solved for its quantum Riemannian geometry in \cite{Ma:sq}. 

\begin{figure}
	\centering
	\begin{tikzpicture}
	\draw (0,0) +(0:1cm) node[right]{$n-1$} -- +(-60:1cm) node[below]{$0$} --  (-120:1cm)node[below]{$1$} (-240:1cm) node[left]{$i$} -- (60:1cm) node[right]{$i+1$};
	\draw (0,0) +  (-120:1cm) -- +(180:1cm)node[left]{$2$};
	\draw [dashed] (0,0) + (180:1cm) --  +(120:1cm);	
	\draw [dashed] (0,0) + (0:1cm) --  +(60:1cm);
	\draw (0,0) +(90:1.2cm) node [above, text width=2.3cm] { $a(i) = g_{i\rightarrow i+1}$ $b(i+1) = g_{i+1\rightarrow i}$}
	          +(270:1.2cm) node [below, text width=2cm] { $a(0) = g_{0\rightarrow 1}$ $b(1) = g_{1\rightarrow 0}$};

	\end{tikzpicture} 
	\caption{A quantum metric on $\Z_n$ is given by metric coefficient functions $a,b$ or equivalently by directed edge weights $g_{i\to i\pm 1}$.\label{figpoly}}
\end{figure}
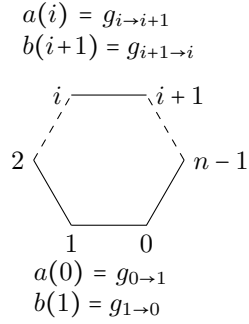

Since the $e^\pm$ are a basis over the algebra, a bimodule invertible quantum metric must take the central form 
\begin{equation} g = a \twoForm{+}{-} + b \twoForm{-}{+}\end{equation}
 for non-vanishing functions $a,b \in \R(\Z_n)$, with inverse metric 
\begin{equation} (e^+,e^+) = (e^-,e^-) = 0, \quad (e^+,e^-) = 1/R_+(b), \quad (e^-,e^+) = 1/R_{-}(a).\end{equation}
 We write $R_\pm=R_{\pm 1}$ for the shift operators.  We also have an inner element $\Theta = e^+ + e^-$ and the canonical $*$-structure $(e^+)^{*} = -e^-; (e^-)^{*} =-e^+$. On the other hand, from the graph perspective, the relevant Cayley graph of $\Z_n$ with the above generators is a polygon of $n$ sides where the values of the functions $a,b$ are directed edge weights according to Figure~\ref{figpoly}. 
From this, it is clear that the edge-symmetric case, where each side of the polygon has weight independent of the direction, requires $b = R_{-}a$. Proceeding in this case, the quantum metric is therefore 
\begin{equation}\label{metricZ} g=ae^+\tens e^-+R_-(a)e^-\tens e^+,\quad (e^+,e^-) = {1\over a}, \quad (e^-,e^+) = {1\over R_{-}a}\end{equation}
as governed by one nonzero function $a$. For convenience, we define  functions on $\Z_n$,
\begin{equation} \rho = {R_{+}(a)\over a}.\end{equation}

\begin{proposition}\label{QLCZn} For $n\ge 3$, there is a  $*$-preserving QLC for any given edge-symmetric  metric (\ref{metricZ}) on $\Omega^1(\Z_n)$. This is the unique for $n\ne 4$ and coincides with the restriction to periodic metrics mod $n$ of the unique such connection on $\Z$ in \cite{Ma:haw}, namely
\begin{align*}
\sigFun{+}{+} = \rho \twoForm{+}{+}, \quad
\sigFun{+}{-} = \twoForm{-}{+}, \\
\sigFun{-}{+} = \twoForm{+}{-}, \quad
\sigFun{-}{-} = R_{-}^2\rho^{-1} \twoForm{-}{-}
\end{align*}
with the geometric structures
\begin{align*}
	\nabla e^+ &= (1-\rho) \twoForm{+}{+}, \quad
	\nabla e^- = (1-R_{-}^2\rho^{-1}) \twoForm{-}{-}, \\
	R_{\nabla}e^+ &= \partial_{-}\rho e^+\wedge \twoForm{-}{+}, \quad
	R_{\nabla}e^- = -\partial_{+}(R_{-}^2\rho^{-1}) e^+\wedge \twoForm{-}{-}, \\
{\rm Ricci}&= \frac{1}{2}\left(\partial_{-}(R_{-}\rho) \twoForm{-}{+} - \partial_{-}\rho^{-1} \twoForm{+}{-} \right), \\
	S &= \frac{1}{2}\left( -\frac{\partial_{-}\rho^{-1}}{a} + \frac{\partial_{-}(R_{-}\rho)}{R_{-}a} \right), \quad
	\Delta f = -\frac{R_{-}{\rho} +1}{a}(\del_++\del_-)f. 
\end{align*}
(For $n=4$, there is a second $*$-preserving QLC given below.)
 \end{proposition}
 \proof Due to the grading restrictions for a bimodule map, the most general $\sigma$ for $n\ne 4$ has the form
\begin{align}
\sigFun{+}{+} = \sigma_0 \twoForm{+}{+} , \quad
\sigFun{+}{-} = \sigma_1 \twoForm{+}{-} + \sigma_2 \twoForm{-}{+}, \nonumber\\ 
\sigFun{-}{+} = \sigma_3 \twoForm{+}{-} + \sigma_4 \twoForm{-}{+}, \quad
\sigFun{-}{-} = \sigma_5 \twoForm{-}{-}
\end{align}
(where the $\sigma_i$ are functional parameters) while for $n=4$ we can have additional terms leading to another solution (given below). Similarly, for $n\ne 3$ we can only have the map $\alpha = 0$ while for $n=3$ we may have additional terms leading to non $*$-preserving solutions in the Appendix. Taking the displayed main form of $\sigma$ and $\alpha=0$,  torsion freeness $\wedge(\id+\sigma)=0$ amounts to 
\begin{equation} \sigma_2 = \sigma_1 + 1,\quad \sigma_3 = \sigma_4 + 1, \end{equation}
while metric compatibility is
\begin{gather}
R_{+}(a) = aR_{+}(\sigma_3)\sigma_0, \quad
a = aR_{+}(\sigma_4)\sigma_1 + R_{-}(a)R_{-}(\sigma_0)\sigma_3, \nonumber\\
R_{-}(a) = aR_{+}(\sigma_5)\sigma_2 + R_{-}(a) R_{-}(\sigma_1)\sigma_4, \quad
R_{-}^2(a) = R_{-}(a)R_{-}(\sigma_2)\sigma_5, \nonumber\\
0 = aR_{ 1}(\sigma_5)\sigma_1 + R_{-}(a)R_{-}(\sigma_1)\sigma_3, \quad
0 = aR_{+}(\sigma_4)\sigma_2 + R_{-}(a)R_{-}(\sigma_0)\sigma_4.
\end{gather}
It is then a matter of solving these, which was done using SAGE\cite{sagemath}. Among the solutions, we find a unique one
that is $*$-preserving. The others are described for completeness in the Appendix. Note that the form of $\Delta$ in comparison to the usual lattice Laplacian makes it clear that $a$ has units of length${}^2$ \cite{Ma:haw,Ma:sq}. \endproof 

That the restriction of the unique $*$-preserving QLC on $\Z$ in \cite{Ma:haw} to periodic metrics  gives a $*$-preserving QLC is not surprising, but that this gives all $*$-preserving solutions for $n\ne 4$ is a nontrivial result of solving the equations as described.  For $n=4$, similar methods lead to a further 2-parameter moduli of  $*$-preserving  connections of the form
\begin{align*}
	\sigFun{+}{+} &=  \gamma \twoForm{-}{-} , 
	\quad \sigFun{+}{-} = - \twoForm{+}{-}, \\
	\sigFun{-}{+}& =  -\twoForm{-}{+}, 
	\quad	\sigFun{-}{-} = {R_+a\over R_-(a\gamma)}  \twoForm{+}{+}, 
\end{align*}
where $\gamma = (\gamma_0, \gamma_1, \bar\gamma^{-1}_0, \bar\gamma^{-1}_1)$ specifies a function on the four points of $\Z_4$ (in order) in terms of 
two complex parameters $\gamma_0,\gamma_1$, such that  $R_+^2\gamma=\bar\gamma^{-1}$. The associated quantum geometric structures are
\begin{align*}\nabla e^+ &=  \twoForm{+}{+} + \twoForm{-}{+} + \twoForm{+}{-} - \gamma\twoForm{-}{-},\\
	\nabla e^- &= \twoForm{-}{-} + \twoForm{+}{-} + \twoForm{-}{+} - r\twoForm{+}{+},\\
	R_\nabla e^+ &= \left( R_-r -1\right) e^+\wedge e^- \tens e^+,\quad R_\nabla e^- = \left(1-  r \right) e^+\wedge e^- \tens e^-,  \\
	{\rm Ricci} &= {1\over 2}\left( R_+r -1 \right) \twoForm{+}{-} + {1\over 2}\left(  R_+^2r-1  \right) \twoForm{-}{+},\\
	S &= {1\over 2a}\left( ({R_-\rho} )(R_+^2r - 1 ) + R_+r -1 \right),\\
	\Delta f&=0,
\end{align*}
where we use the shorthand
\begin{equation} r:= {R_+(a)\over R_-(a)}|\gamma|^2.\end{equation}
This is the $*$-preserving case of the general $n=4$ solution (i)  in the Appendix. 

\subsection{The circle limit of the $\Z_n$ quantum geometry} \label{seccirc}

We now turn to the matter of the classical limit $n\to\infty$ of the quantum geometry on $\Z_n$. Given that $\Omega^1(\Z_n)$ is 2-dimensional, we can not expect exactly a classical circle in the limit. 

To put the quantum geometry in a more convenient form, we first (Fourier) transform to a new variable $s$, where $s\in \C(\Z_n)$ is defined by
\begin{equation}\label{sq} s(i)=q^i,\quad q=e^{2\pi\imath\over n},\quad \C(\Z_n)\cong\C\Z_n:=\C[s]/(s^n-1)\end{equation}
In this new description, our same algebra $A$ is generated by $s$ with the relation $s^{n}=1$. Also note that
\begin{equation}\extd s^{-1}=- s^{-1}(\extd s)s^{-1}\end{equation}
is independent of $\extd s$ until we specify the commutation relations of $\extd s$ with $s$. We thus define two left-invariant 1-forms
\begin{equation} f^+:=s^{-1}\extd s,\quad f^-:=s\extd s^{-1}.\end{equation}
For the $n\to \infty$ limit, we can now just drop the $s^n=1$ relation so that $A=\C[s,s^{-1}]$, the algebraic circle with $s^*=s^{-1}$. On can think of this as $s=e^{\imath\theta}$ in terms of an angle coordinate $\theta$. Its {\em classical} differential calculus has $\extd s$ central and hence one left-invariant 1-form $\bar f^+=\imath\extd\theta=-\bar f^-$, and the standard constant metric is 
\begin{equation} \extd\theta\tens\extd\theta= -\bar f^+\tens \bar f^+.\end{equation} We are not in this classical case. We set $[m]_q:=(1-q^m)/(1-q)$ as the usual $q$-deformed integer. 

\begin{proposition}\label{qcircle}  In these new coordinates, the $f^\pm$ form a Grassmann algebra and 
\[  f^-s=-s f^+,\quad f^+s=s(f^-+(q+q^{-1}) f^+),\quad  \extd s^m=-{q[m]_q s^m\over (q+1)}\left(q[-1-m]_qf^++ [1-m]_qf^-\right),\]
while the $*$-operation and the element that makes the calculus inner are 
\[ f^\pm{}^*=-f^\pm,\quad \Theta={q\over (q-1)^2}\Theta_0;\quad \Theta_0=f^++f^-\]
and the constant $a=1$ metric $g=e^+\tens e^-+e^-\tens e^+$ is
\[ g={g_0 \over (q-q^{-1})^2};\quad g_0= -2 f^+\tens f^++\Theta_0\tens f^++f^+\tens\Theta_0+ {2q\over (q-1)^2}\Theta_0\tens\Theta_0.\]
Moreover, the above does not require $s^n=1$, i.e. applies equally well to the algebraic circle $\C[s,s^{-1}]$ with $q$ a real or modulus 1 free parameter. 
\end{proposition}
\proof Working in our original calculus $\Omega(\Z_n)$ and $s,q$ the function and the root of unity specified in (\ref{sq}), we compute that
\begin{equation} f^-=s\extd s^{-1}=(q^{-1}-1)e^++ (q-1)e^-,\quad f^+=s^{-1}\extd s=(q-1)e^++(q^{-1}-1)e^-\end{equation}
which inverts for $n>2$ as 
\begin{equation} e^+={f^-+q f^+\over (q-q^{-1})(q-1)},\quad e^-={f^++q f^-\over (q-q^{-1})(q-1)}.\end{equation}
As they are linear combinations, the $f^\pm$ are closed and form a Grassmann algebra since the $e^\pm$ do. We have $e^\pm s= R_\pm(s)e^\pm=q^{\pm 1}s e^\pm$ which implies the relations shown for $f^\pm$. Finally, $\extd s^m=(\del_+ s^m)e^++(\del_- s^m)e^-=(q^m-1)e^++ (q^{-m}-1)e^-$ which translates to the formula shown in terms of $f^\pm$. The $*$ structure also matches but is in any case required by $f^+{}^*=(s^{-1}\extd s)^*=(\extd s^{-1})s=s^{-1}f^-s=-f^+$ and similary for $f^-$. We also have $\Theta=e^++e^-$ and $g$ as stated when written as above in terms of $f^\pm$. The quantum Levi-Civita connection now appears equivalently as  $\nabla f^\pm=0$. 

Moreover, these formulae do not directly reference $n$ and one can check directly that they give a $*$-differential calculus even without the relation $s^n=1$, i.e. on the algebraic circle. Now $q$ is a free parameter but a check shows that we still need it real or modulus one for a $*$-calculus. 
\endproof

The end result of Proposition~\ref{qcircle} is a novel, 2-dimensional, $q$-deformed calculus on the algebraic circle. In the $q$ real case, we can quotient it by a relation such as  $f^+=-qf^-$, which is equivalent to the relation $e^-=0$ and gives the standard 1-dimensional $q$-deformed calculus on the circle \cite[Ex 1.11]{BegMa:book} with $\extd s.s=q s\extd s$ or $\extd s^m=[m]_q s^{m-1}\extd s$. In this quotient, we would have  $g=0$ (this quotient calculus in fact admits no quantum metric due to the centrality requirement,  making it unsuitable for our purposes).   

\begin{corollary}\label{circlim} In the limit $q\to 1$, the above $q$-deformed calculus on the circle algebra $\C[s,s^{-1}]$ tends to a noncommutative 2D calculus with 
\[ f^-s=-s f^+,\quad f^+s=s(f^-+2 f^+),\quad\extd s^m={m s^m\over 2}\left((m+1)f^++ (m-1)f^-\right), \quad f^\pm{}^*=-f^\pm\]
In this limit, the 1-form $\Theta_0$ is closed and graded-central and the classical calculus on $S^1$ is then given by the quotient where we set  $\Theta_0=0$. Conversely, this 2D calculus is a central extension in the sense of \cite{Ma:alm,Ma:rec} of the classical calculus on $S^1$ by $\Theta_0$.  
\end{corollary}
\proof Most of this is immediate. For the last sentence, note that if $f$ is a function of $s$ then we can write the differential in the corollary equivalently as 
\begin{equation}\label{lapcirc} \extd f= s{\extd f\over\extd s}f^++ {s^2\over 2}{\extd^2 f\over\extd s^2}\Theta_0,\end{equation}
where the first term is the expected left-invariant derivative associated to $f^+$ and the second is a higher order derivative associated to an `extra direction' $\Theta_0$. This has the structure of a central extension of the classical calculus on $S^1$ in the sense of \cite[Prop.~1.22]{BegMa:book}\cite{Ma:alm,Ma:rec} according to the canonical Riemannian structure of $S^1$ and a second order operator with respect to it. The central extension here is defined by a deformed $\bullet$ product where $s\bullet \bar f^+=s\bar f^+$ is undeformed for left multiplication on the classical left-invariant 1-form $\bar f^+=s^{-1}\extd s=\imath\extd\theta$. From the other side, we set $\bar f^+\bullet s=\bar f^+ s+ (\bar f^+,\extd s)\Theta_0= s\bar f^++s(\bar f^+,\bar f^+)\Theta_0=s\bullet \bar f^+ + s\Theta_0$, which is the stated commutation relation for $f^+$ if we take the classical constant metric on $S^1$ with normalisation $(\bar f^+,\bar f^+)=1$. As $\Theta_0$ commutes with functions, this determines the correct commutation relation for $f^-$ also.  The second order operator defines $\extd$ and here is $s^2{\extd^2\over\extd s^2}$, which is the Laplacian plus a vector field as an example of the general set up \cite[Thm~8.23]{BegMa:book}\cite{Ma:alm}.  \endproof

Next, we note that the rescaled metric $g_0$ in Proposition~\ref{qcircle} has a part with a $q\to 1$ limit plus a singular term proportional to $\Theta_0\tens\Theta_0$. Hence, if $\pi_{class}$ denotes taking $q\to 1$ and simultaneously projecting to the classical calculus, we have
\begin{equation} \pi_{class}(g_0)= -2\bar f^+\tens \bar f^+=2\extd\theta\tens\extd\theta,\end{equation}
{\em provided that} in this process, the killing of $\Theta_0$ takes precedence over setting $q\to 1$ in the singular term. It is not clear how to make this precise (one cannot simply set $\Theta_0=0$ first without destroying the structure of the $q$-deformed calculus). Aside from this technical detail, we still have  the trivial flat QLC  $\nabla f^\pm=0$ and the projection is covariantly constant with respect to this and the usual classical connection. We have focussed here on the limit of the constant metric on $\Z_n$, but one can analyse general metrics in the similar way. Also, in the $q\to 1$ limit as in   Corollary~\ref{circlim}, one can directly analyse the possible generalised (not necessarily quantum-symmetric) metrics, e.g. the ones with constant coefficients have the form 
\begin{equation} g={\rm Re}(z) (f^+\tens f^++f^-\tens f^-)+ z f^+\tens f^-+ \bar z f^-\tens f^+\end{equation} 
for a complex parameter $z$ in order to be central and obey the reality condition. If we then impose quantum symmetry, we are forced to a real multiple of $\Theta_0\tens\Theta_0$, which is indeed the only component of the flat metric $g$ if we fully scale out the singularity visible in Proposition~\ref{qcircle}  and then set $q\to 1$.  This is a `purely quantum' metric in the 2D calculus in Corollary~\ref{circlim}, in that it projects to zero in the classical calculus on $S^1$. 

 We have already seen that the extra direction $\Theta_0$ of the calculus in Corollary~\ref{circlim} arises as the residue of the element $\Theta$ that makes the $q$-deformed calculus on the circle inner, which is a purely quantum phenomenon. It can also be viewed as defining a central extension of the classical calculus on $S^1$ with associated `partial derivative' the second order operator in (\ref{lapcirc}). A third point of view is given by moving to `cartesian coordinates' 
\begin{equation} x={1\over 2}(s+s^{-1}),\quad y={1\over 2\imath}(s-s^{-1});\quad x^2+y^2=1\end{equation}
from which we compute
\begin{equation}  2(x\extd x+ y\extd y)=s^{-1}\extd s+s\extd s^{-1} =\Theta_0.\end{equation}
Thus, $\Theta_0$ can be thought of as something like the normal to the circle viewed in the plane, similarly to the picture for the extra direction for the 3D calculus on the fuzzy sphere in \cite{LirMa}.

Finally, in cohomological terms, one can check that the  noncommutative de Rham cohomology ring $H_{\rm dR}(\Z_n)$ is the Grassmann algebra generated by $e^\pm$ i.e. dimensions $1:2:1$ and spanned by $e^\pm$  in degree 1. The same is true in terms of the $f^\pm$ for finite $n$, which  latter description holds also for $n\to \infty$; $H_{\rm dR}(\Z)$ is generated by $f^\pm$ in the case of the corollary. This is the same as the cohomology of a torus, so it is tempting to think of the quantum geometry as a circle thickened into a torus, at least in a  cohomological sense. The geometric picture, as we have seen, is a little like this with an extra direction related to the normal to the circle (rather than an actual torus).

\subsection{Euclideanised quantum gravity on $\Z_n$}\label{secqg}

As for the integer line graph\cite{Ma:haw}, the two-dimensional cotangent bundle on $\Z_n$ required by the quantum geometry now admits the possibility of curvature. We envision that there could be various applications of such curved discrete geometries, but here we focus on just one, namely Euclideanised quantum gravity on $\Z_n$. For integration on $\Z_n$ needed in the action, we take a sum over $\Z_n$ with a weight $a$ (in the commutative case, this would be $\sqrt{|\det g|}$), which has the merit that then the action is 
\begin{equation} S_g={1\over 2}\sum_{\Z_n}(R_-\rho\del_- R_-\rho)={1\over 2}\sum_{\Z_n}\rho\del_-\rho={1\over 2}\sum_{\Z_n}\rho\del_+\rho={1\over 4}\sum_{\Z_n}\rho(\del_++\del_-)\rho, \end{equation}
where $\del_++\del_-$ is the usual lattice double-differential on $\Z_n$. This has the same form as for a scalar field except that $\rho$ is a positive function, as already observed for $\Z$ in \cite{Ma:haw}. We consider two approaches, depending on 
what we regard as our underlying field, and in both cases maintaining  $\Z_n$ symmetry in the result. 

 (i) As suggested by the form of the action, we can take
\begin{equation} \rho_0={a(1)\over a(0)},\quad \cdots,\quad\rho_{n-2}={a(n-1)\over a(n-2)},\quad \rho_{n-1}={a(0)\over a(n-1)}\end{equation}
as $n$ dynamical variables subject to the constraint $\rho_0\cdots\rho_{n-1}=1$. We think of the constraint as a hypersurface in $\R_{>0}^n$, which induces a metric $\cg_\rho$ on the hypersurface, and we use the Riemannian measure in this. Thus, we can take $\rho_0,\cdots,\rho_{n-2}$ as local coordinates and measure ${\mathcal D}\rho=(\prod_{i=0}^{n-2}\extd\rho_i)\sqrt{\det(\cg_\rho)}$. The measure here maintains the $\Z_n$ symmetry as ultimately independent of the choice of coordinates. 

Explicitly, for $n=3$, we take $\rho_0,\rho_1$ as coordinates and the constrained surface in $\R_{>0}^3$ is $\rho_2=1/(\rho_0 \rho_1)$. The coordinate tangent vectors and induced metric are
\begin{equation} v_0=(1,0, -{1\over\rho_0^2\rho_1}),\quad v_1=(0,1,-{1\over\rho_0\rho_1^2});\end{equation}
\begin{equation} \cg_\rho=(v_i\cdot v_j)=\begin{pmatrix} 1+{1\over\rho_0^4\rho_2}& {1\over\rho_0^3\rho_1^3}\\ {1\over\rho_0^3\rho_1^3}& 1+ {1\over\rho_0^2\rho_1^4}\end{pmatrix},\quad \det(\cg_\rho)=1+{1\over\rho_0^4\rho_1^2}+{1\over\rho_0^2\rho_1^4}.\end{equation}
Hence the partition function is 
\begin{equation} Z= \int_0^\infty\extd\rho_0\int_0^\infty\extd\rho_1\sqrt{\det(\cg_\rho)}\, e^{- {1\over 2 \Gn }(\rho_0^2+\rho_1^2+\rho_2^2-\rho_0\rho_1-\rho_1\rho_2-\rho_2\rho_0)};\quad \rho_2:={1\over\rho_0\rho_1}\end{equation}
These integrals can be done numerically and appear to converge for all values $\Gn >0$ of the coupling constant (the numerical results need $\Gn $ not too small for working precision but this case can be analysed separately). We are interested in expectation values $\<\rho_{i_1}\cdots\rho_{i_m}\>$, where we insert $\rho_{i_1}\cdots\rho_{i_m}$ in the integrand and take the ratio with $Z$. 

\begin{figure}
\[ \includegraphics[scale=0.7]{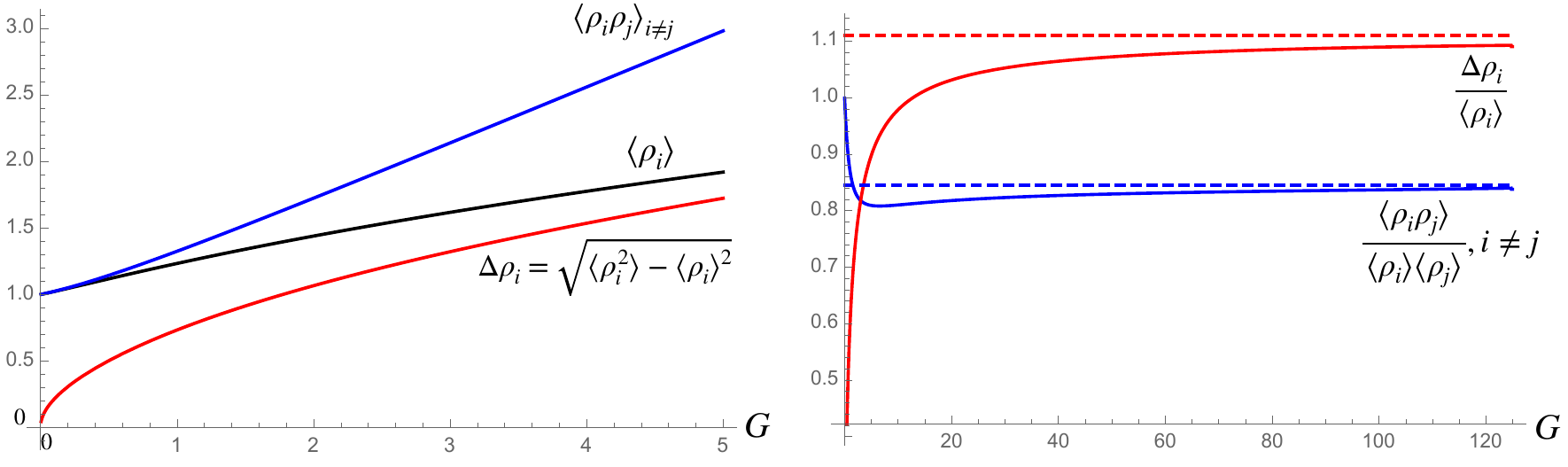}\]
\caption{Euclidean quantum gravity vevs on $\Z_3$ for gauge invariant variables $\rho_i$\label{figrho}}
\end{figure}

\begin{figure}
\[ \includegraphics[scale=0.9]{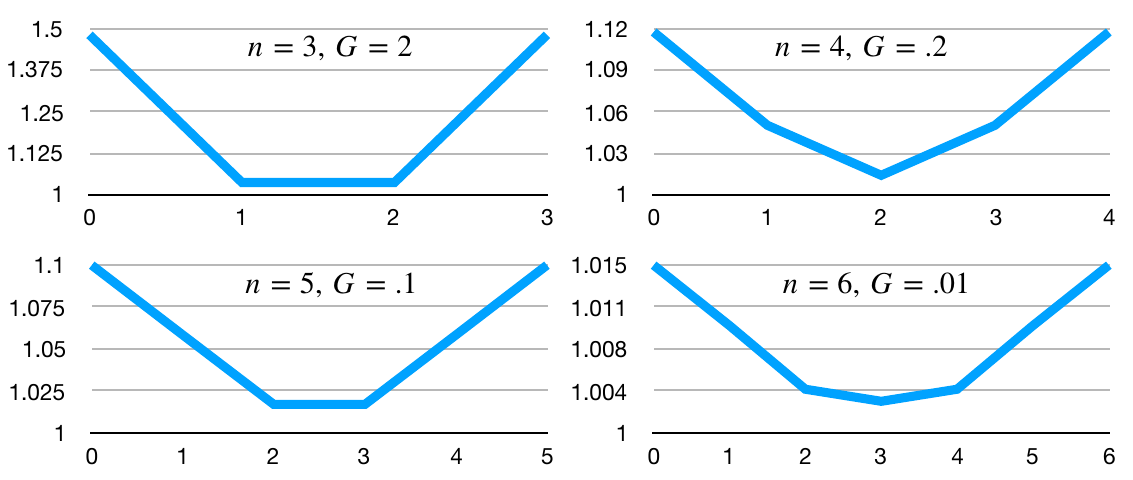}\]
\caption{Euclidean quantum gravity correlations  $\<b_0b_i\>$ plotted against $i$ for $3\le n\le 6$ and suitable $\Gn $. \label{figZn}}
\end{figure}

Some results obtained from this theory for $n=3$ are plotted in Figure~\ref{figrho}. Numerical evidence is limited due to convergence accuracy issues, but it seems clear that expectation values of products of $\rho_i$ tend to 1 and hence $\Delta \rho_i\to 0$ as $\Gn \to 0$, as might be expected. As in \cite{Ma:sq}, this should be thought of as the weak gravity limit given that fluctuations expressed in $\rho$ enter the action relative to $\Gn $.  Meanwhile,  it appears as $\Gn \to \infty$ that 
\begin{equation} {\Delta \rho_i\over \<\rho_i\>}\sim 1.11,\quad {\<\rho_i^2\>\over\<\rho_i\>^2}\sim 2.23,\quad {\<\rho_i\rho_j\>\over \<\rho_i\>\<\rho_j\>}\sim 0.845\end{equation}
for $i\ne j$. The asymptotic values here are from plotting out to $\Gn=500$, but would need to be confirmed analytically
due to potential numerical convergence issues. The first of these limits, if confirmed, would be a similar phenomenon of a uniform the relative metric uncertainty in \cite{Ma:sq} in the `strong gravity' limit.  The correlations are real and relative correlation between two distinct vertices of the triangle is lower than the relative self-correlation, which is in line with the $n=3$ case of  the relative quantisation in Figure~\ref{figZn}. 

(ii) We can take (as more usual) the metric coefficients as the underlying field, so in our case the edge `square-lengths'  $a=(a_0,\cdots,a_{n-1})\in\R_{>0}^n$. Assuming  Lebesgue measure,  the partition function is
\begin{equation} Z=\int_0^\infty (\prod_i\extd a_i)  e^{{S_g\over \Gn }}=\int_0^L (\prod_i\extd a_i)  e^{ {1\over 2 \Gn }\sum_{\Z_n}\rho\del_+\rho}\end{equation}
 and we introduce a field strength upper bound $L$ to control divergences as in \cite{Ma:sq}. One can then look at ratios independent of $L$ or indeed consider a formal renormalisation process. 
 
On the other hand, the divergences come from the global scaling symmetry $a_i\mapsto \lambda a_i$ for $\lambda\in \R_{>0}$ of the action (since this depends only on the ratios $\rho$) and therefore another approach would be to `factor out' the geometric mean as a new variable which we do not integrate over, keeping only the ratios relative to this as the dynamic degrees of freedom. This is again in the spirit of \cite{Ma:sq}, except that we proceed multiplicatively. Thus, we let $A=(\prod_i a_i)^{1\over n}$ be the geometric mean and $b_i:=a_i/A$, which by construction obey $b_0\cdots b_{n-1}=1$. These are similar to the $\rho_i$ variables in forming the corresponding hypersurface in $\R_{>0}^n$, but the action is different and the measure is also different since it is inherited from the Lebesgue measure on the $a_i$.
 
 Again, we will look at this explicitly for $n=3$. Then the action is 
 \begin{equation} S_g={1\over 2}\left({b_0\over b_1}+{b_1\over b_2}+ {b_2\over b_0}- ({b_1\over b_0})^2-({b_2\over b_1})^2-({b_0\over b_2})^2\right);\quad b_2={1\over b_0 b_1},\end{equation}
 while the Jacobean for the change of variables from $a_0,a_1,a_2$ to $b_0,b_1,A$ gives us
 \begin{equation} \extd a_0\, \extd a_1\, \extd a_2={3 A^2\over b_0 b_1}\extd b_0\, \extd b_1\, \extd A.\end{equation}
 Omitting  the now decoupled integration over $A$ as an (infinite) constant, we have
 effectively 
 \begin{equation} Z= \int_0^\infty\extd b_0\int_0^\infty\extd b_1{1\over b_0 b_1} e^{ {1\over 2 \Gn  b_0^2 b_1^4}(-1+(1+b_0^3)b_1^3+( -1+b_0^3-b_0^6) b_1^6)}.\end{equation} 
 The graphical expectation values against $\Gn $ look qualitatively similar to those of $\rho_i$ in Figure~\ref{figrho}, but one also has $\<b_i\>=\<b_i b_j\>$ for $i\ne j$, albeit this is specific to $n=3$. 
 
 Larger $n>3$ can proceed entirely similarly and one has $1<\<b_i\>< \<b_i b_{i+1}\>$. One can also then see that the $i$-step correlations  $\<b_0b_i\>$ (or between any two points differing by $i$) decrease as $i$ increases from $i=0$ to reach a minimum (as expected) half way around the polygon. This is based on numerical data for small $n$ as shown in Figure~\ref{figZn}. The data for $n=6$ are already noisy due to numerical convergence issues, but suggest that for large $n$ the $\<b_0b_i\>$ may be approximated by $\alpha-\beta\sin({\pi i\over n})$ for positive $\alpha>\beta$ depending on $G$ and $n$. This is broadly similar to the form of correlation functions for a scalar field $\<\phi_0\phi_i\>$ in a lattice box in \cite{Ma:haw}, but without the overall $\imath$ there.

\section{Quantum geometric cosmological models on $\R\times\Z_n$}\label{secRpoly}

In this section, we first start with an analysis of quantum metrics and QLCs on $\R\times\Z_n$, where $\R$ is a classical time and $\Z_n$ is a discrete space. We find that the full `strongly tensorial' bimodule properties for  an invertible quantum metric force us to the block diagonal case, without taking this as an assumption. Existence of a QLC further dictates its form, again without taking this as an assumption, and we then find a unique $*$-preserving one.  We then focus on the case where the $\Z_n$ geometry is flat (modelling an actual geometric circle) but possibly time-dependent as in FLRW cosmology.

\subsection{Quantum metric and QLC on $\R\times\Z_n$} \label{seccross}

We consider a general metric on the product $\mathbb R\times \Gp $ where $\R$ has a variable $t$ and we are interested in the finite group $\Gp =\Z_n$ with $e^a=e^\pm$, but we do not need to specialise at this stage. We consider metrics of the form
\begin{equation}g=\mu\extd t\otimes\extd t+ h_{ab}e^a\otimes e^b+ n_a(e^a\otimes\extd t+\extd t\otimes e^a)\end{equation}
for $\mu, h_{ab}, n_a$ in $A=\C^\infty(\R)\otimes \mathbb C(\Gp )$ but note right away that if we take the tensor product calculus where the continuous variable and its differential $t,\extd t$ graded commute with functions and forms on $\Gp $ then centrality of the metric needed for a bimodule inverse dictates that $n_a=0$. We therefore proceed in this case. 

Similarly, we look for general QLCs of the form
\begin{equation} \nabla \extd t=-\Gamma \extd t\tens\extd t+ c_a(e^a\tens\extd t+\extd t\tens e^a)+ d_{ab}e^a\tens e^b,\end{equation}
\begin{equation} \nabla e^a=-\Gamma^a{}_{bc}e^b\tens e^c+ \gamma^a{}_b(e^b\tens\extd t+\extd t\tens e^b)+ f^a\extd t\tens\extd t\end{equation}
and note that for the tensor form of calculus along with the natural choice where $\sigma(\extd t\tens \ ),\sigma(\ \tens\extd t)$ are the flip on the basic 1-forms $\extd t, e^a$, requiring the above to be a bimodule connection compatible with the relations of each algebra forces us to
\begin{equation} c_a=0,\quad f^a=0,\quad  \gamma^a{}_b=\gamma_a\delta_{a,b},\quad d_{a,b}=d_a\delta_{a,b^{-1}}\end{equation}
for some functions $\gamma_a$. We therefore proceed in this case. 

Next,  for zero torsion, we need that 
\begin{equation} d_{ab}=d_{ba},\quad \Gamma^a{}_{bc}=\Gamma^a{}_{cb},\quad \wedge(\id+\sigma)(e^a\tens e^b)=0\end{equation}
(which means $\sigma$ restricted to the $\{e^a\}$ has the form studied before for a torsion free bimodule connection on an inner calculus, but note the calculus as a whole is {\em not}  inner). And for  $\nabla g=0$, we obtain  8 equations which we compute under our assumptions above for a central metric and bimodule connection, with $\dot\mu={\partial\over\partial t}\mu$, 
\begin{align}\extd t^{\tens 3}:&\quad {\dot\mu\over 2}-\mu\Gamma = 0, \\
\extd t\tens\extd t\tens e^a:&\quad 0=0,\\
 \extd t\tens e^a\tens\extd t:&\quad0=0,\\
 e^a\tens \extd t\tens\extd t:&\quad \del_a\mu=0,
 \\
 \extd t\tens e^a\tens e^b:&\quad   h_{cb}\gamma^c{}_a+h_{ac}R_a(\gamma^c{}_b)+ \dot{h}_{ab} =0,\\
 e^a\tens \extd t\tens e^b:&\quad   h_{cb}\gamma^c{}_a+\mu d_{ab} =0, \\
  e^a\tens e^b\tens   \extd t:&\quad  \mu d_{ab}+h_{mp}R_m(\gamma^p{}_n)\sigma^{mn}{}_{ab}=0, \\
 e^m\tens e^n\tens e^p:&\quad \del_m h_{np}-h_{ap}\Gamma^a{}_{mn}-h_{ac}R_a(\Gamma^c{}_{bp})\sigma^{ab}{}_{mn}=0. 
 \end{align}
 The first and last of the 8 equations  are just that $\Gamma$ is a QLC on the line and $\sigma,\Gamma^a{}_{bc}$ a QLC on $\Gp$. The 4th equation tells us that $\mu$ is constant on $\Gp $. If we write the metric as $h_{ab}=h_a\delta_{a,b^{-1}}$ for functions $h_a$ etc., then the 6th equation tells us 
\begin{equation}\label{fcd}  d_a=-{h_{a}\gamma_{a}\over \mu}\end{equation}
and the 5th and 7th equations reduce to
\begin{equation}\label{gamma} \dot h_a+h_a\gamma_a+R_a(h_{a^{-1}}\gamma_{a^{-1}})=0,\quad \sum_p R_{p^{-1}}(h_p\gamma_p)\sigma^{p^{-1},p}{}_{a,b}=h_a\gamma_a\delta_{a,b^{-1}}.\end{equation}
Finally, we impose $*$-structure $\extd t^*=\extd t$ and suppose that the connection on $\Gp $ is also $*$-preserving for $e^a{}^*=-e^{a^{-1}}$ as usual. The extended metric then obeys the quantum reality condition if $\mu$ is real, which we  suppose henceforth,  and the metric on $\Gp $ is `real' in the required sense (which amounts to $h_a$ real-valued). Then the additional condition for our extended $\nabla$ to be $*$-preserving comes down to $\Gamma$ real  and 
\begin{equation}  \bar\gamma_a=R_a\gamma_{a^{-1}}, \quad \sum_a \bar d_a\sigma(e^a\tens e^{a^{-1}})=\sum_a d_{a^{-1}}e^{a^{-1}}\tens e^a, \end{equation}
where the 1st part comes from $\nabla e^a{}^*$ and the 2nd from $\nabla\extd t^*$. Next, we use (\ref{fcd}) and that $h_a$ are real and edge-symmetric to deduce from the 1st part that $\bar d_a=R_a d_{a^{-1}}$. Then since $d_a$ are constant on $\Gp $, we have $\bar d_a=d_{a^{-1}}$ and our condition to be $*$-preserving is 
\begin{equation}\label{starpres}  \bar\gamma_a=R_a\gamma_{a^{-1}}, \quad \sum_a d_{a^{-1}}(\sigma(e^a\tens e^{a^{-1}})-e^{a^{-1}}\tens e^a)=0.\end{equation}

Since $\mu$ has to be a constant on $\Gp $, it is some function of $t$ alone. Generically, we can absorb this in a change of the variable $t$, so we proceed for simplicity with $\mu=-1$ for a cosmological type solution. 

\begin{theorem} \label{thm:connection} For $\sigma,\nabla^{\Z_n}$ the $*$-preserving QLC on $\Z_n$ in Propostion~\ref{QLCZn}, a quantum metric on $\R\times\Z_n$ admitting a $*$-preserving QLC has the form
\[g=-\extd t\otimes\extd t- a e^+\tens e^--R_-a e^-\tens e^+\]
up to choice of the $t$ parametrization, such that $\del_-\dot a=0$, i.e.,  $a$ has the form
\[ a(t,i) = \alpha (t) + \beta(i)\]
for some functions $\alpha,\beta$ with $\sum_i\beta(i)=0$. In these terms, there is a  unique $*$-preserving QLC with scalar curvature and Laplacian
\begin{align*}
	2S =&  - \ddot{\alpha}\left(\frac{ 1}{\alpha  + \beta}  + \frac{1}{\alpha  + R_-\beta} \right) + \frac{\dot{\alpha}^2}{4} \left( \frac{1}{(\alpha  + \beta)^2} + \frac{1}{(\alpha  + R_-\beta)^2} \right)        \\
	&+ \frac{  s  }{    (\alpha+\beta)^2(\alpha+R_+\beta)  } + R_-\left(\frac{ s  }{(\alpha+\beta)^2(\alpha+R_-\beta)}\right),\\
	\Delta f=&-\del_t^2+\left({1\over \alpha+\beta}+{1\over \alpha+R_-\beta}\right)(-{\dot\alpha\over 2}\del_t f+ \Delta_{\Z_n}f), 
\end{align*}
where 
\[s:= (\alpha+R_+\beta)(\alpha+R_-\beta)-(\alpha+\beta)^2=\alpha(\Delta_{\Z_n}\beta)+(\del_+\beta)\del_-\beta  - \beta^2 \]
in terms of the usual Laplacian $\Delta_{\Z_n}\beta=(\del_++\del_-)\beta=R_+\beta  + R_-\beta-2\beta$ on $\Z_n$. 
 \end{theorem}
 \proof  We use the general analysis above applied in the specific case of $\Z_n$. Also, for the purpose of the proof, it is convenient to have a shorthand notation $a_+=a$ and $a_-=R_-a$, so that $h_\pm=a_\pm$ for our particular metric. Then the 2nd  of (\ref{gamma}) holds automatically as $\sigma(e^\pm\tens e^\mp)=e^\mp\tens e^\pm$ and   $a_\pm\gamma_\pm=d_\pm(t)$ are constants on $\Z_n$ for a solution, while the 1st of (\ref{gamma}) is that   
$\dot a_\pm=- d_+- d_-$, which requires $\del_-\dot a=0$ as stated. We assume the QLC on $\Z_n$ at each $t$  for the metric functions $a=a(t,i)$.  The flip form of $\sigma(e^\pm\tens e^\mp)$ for this also means that the 2nd part of  (\ref{starpres}) is automatic and we just need $\bar\gamma_\pm=R_\pm\gamma_{\mp}$, or equivalently $\bar d_\pm=d_\mp$, for a $*$-preserving connection. This means that 
\begin{equation} d_+= -{\dot a\over 2} + \imath b,\quad d_-=\bar d_+=-{\dot a\over 2}- \imath b;\quad \gamma_\pm = -{\dot a\over 2 a_\pm} \pm {\imath b\over a_\pm}\end{equation}
for any real-valued function $b(t)$. The unique solution with real coefficients for $\nabla$ in our basis is $b=0$ and gives the $*$-preserving QLC
\begin{equation}\label{QLCRZn} \nabla \extd t = {\dot a\over 2}(e^+\tens e^-+e^-\tens e^+),\quad \nabla e^\pm=\nabla^{\Z_n}e^\pm-  {\dot a\over 2 a_\pm}(e^\pm\tens\extd t+\extd t\tens e^\pm). \end{equation}
The $\sigma$ for this when one argument is $\extd t$ is the flip. We then proceed to compute the curvature  of this QLC, 
\begin{align*}R_\nabla e^\pm&= R^{\Z_n}_\nabla e^\pm- \left(\dot{\Gamma}^\pm{}_{ab} - \Gamma^\pm{}_{ab}R_a({\dot{a}\over2a_b}) + {\dot{a}\over2a_\pm{}}\Gamma^\pm{}_{ab} \right)\extd t\wedge e^a\tens e^b  - \Gamma^\pm{}_{ab}R_a({\dot{a}\over2a_b}) e^a\wedge e^b\tens \extd t\\
&\quad\pm ({\dot{a}\over 2 a_\pm})^2 a_\pm e^+\wedge e^-\tens e^\pm-{\dot a\over 2}\del_b({1\over a_\pm})e^b\wedge e^\pm\tens\extd t + {\dot{a}\over 2}\del_b({1\over a_\pm})\extd t\wedge e^b\tens e^\pm  \\
&\quad - \left( { \del \over \del t}({\dot{a}\over2a_\pm{}}) + ({\dot{a}\over2a_\pm{}})^2 \right)\extd t \wedge e^\pm{} \tens \extd t, \\
 R_\nabla\extd t&={\ddot a\over 2}\extd t\wedge(e^+\tens e^-+e^-\tens e^+)+{\dot a\over 2}e^+\wedge \Gamma^-{}_{-b}e^-\tens e^b+{\dot a\over 2}e^-\wedge\Gamma^+{}_{+b}e^+\tens e^b\\
 &\quad+\sum_\pm({\dot a\over 2a_\pm})^2a_{\pm}e^\pm \wedge (e^\mp\tens\extd t+\extd t\tens e^\mp),\end{align*}
in terms of the Christoffel symbols on $\Z_n$. The Ricci tensor and the Ricci scalar $S$ are then
\begin{align*}
	{\rm Ricci} =&\ {\rm Ricci}^{\Z_n} + {\ddot{a}\over 4} (\twoForm{+}{-} + \twoForm{-}{+}) + {1\over 2}\left( R_+(\dot{\Gamma}^-_{--}) -{\dot{a}\over 2} (R_+(\Gamma^-_{--}) + 1) \del_-\left( {1\over a} \right) \right) \extd t \tens e^- \\ 
	&+ {1\over 2}\left( R_-(\dot{\Gamma}^+_{++})  - {\dot{a}\over 2}(R_-(\Gamma^+_{++}) +1)\del_+\left({1\over a_-}\right) \right) \extd t \tens e^+ + {\dot{a}\over 4}\left( (R_-(\Gamma^+_{+-}) + 1)\del_-\left({1\over a_-}\right)   \right) e^-\tens \extd t \\
	&- {\dot{a}\over 4}\left( (R_+(\Gamma^-_{+-}) + 1)\del_-\left( {1\over R_+(a)} \right)  \right) e^+\tens \extd t + {1\over 2}\left(  \del_t\left( {\dot{a}\over 2a} + {\dot{a}\over 2a_-} \right) + \left({\dot{a}\over 2a}\right)^2 + \left({\dot{a}\over 2a_-}\right)^2  \right)\extd t \tens \extd t, \\
	S =& -S^{\Z_n} - {\ddot{a}\over2}\left( {1\over a} + {1\over a_-} \right) +  {1\over2} \left({\dot{a}\over 2a}\right)^2 + {1\over2}\left({\dot{a} \over 2a_-}\right)^2
\end{align*}
 (where we have used that $\Gamma^\pm_{+-} = \Gamma^{\pm}_{-+}$). We now insert values for the QLC in Proposition~\ref{QLCZn} to obtain
\begin{align}
R_\nabla e^\pm&= \pm\left( -\del_\pm\left({a_\pm\over a_\mp}\right) +  \left( {\dot{a}\over 2a_{\pm}}\right)^{2}a_\pm \right)  e^+ \wedge \twoForm{-}{\pm} 
	+ {\dot{a}\over 2 a^2_\pm}\del_\pm\left(a_\pm \right) \extd t\wedge \twoForm{\pm}{\pm}  \nonumber \\
	&\quad+{ \dot{a}\over 2} \del_\mp\left({1\over a_\pm} \right)( e^\pm\wedge e^\mp\tens \extd t  + \extd t\wedge\twoForm{\mp}{\pm}) \nonumber\\
	&\quad+ \left( -{ \ddot{a} \over 2a_\pm}   + \left( { \dot{a} \over 2a_\pm} \right)^2  \right) \extd t \wedge e^\pm \tens \extd t,  \\
	R_\nabla \extd t&= \sum_{\pm} \left( {\ddot{a}\over 2a_\pm} - \left( \dot{a}\over 2a_\pm \right)^2 \right)a_\pm \extd t \wedge \twoForm{\pm}{\mp} + \sum_\pm {\dot{a}\over 2a_\pm }\del_-( a) e^+ \wedge \twoForm{-}{\mp}\nonumber\\ &\quad+ {\dot{a}^2\over 4} \del_-\left( 1\over a^2 \right)  e^+ \wedge e^- \tens \extd t\end{align}
	and as a result,
 \begin{align}{\rm Ricci}&=   {1\over 2}\sum_\pm \left(\big(  { \ddot{a } \over 2} +\del_\pm\big({a_\mp\over a_\pm }\big) \big)\twoForm{\pm}{\mp} - {\dot{a}\over 2a^2_\mp}\del_\pm(a_\mp) \extd t \tens e^\pm   + {\dot{a}\over 2}\del_\pm \big( {1\over a_\pm } \big) e^\pm \tens \extd t  \right)  \nonumber\\
	&\quad - {1\over 2} \left(-{\ddot{a}\over 2}  \left(  {1 \over a } + {1\over a_-}  \right) + \left( {\dot{a}\over 2a } \right)^2 +\left( {\dot{a}\over 2a_-} \right)^2 \right) \extd t \tens \extd t, 
	\\
	S &={1\over2}\left(-   \ddot{a}     \left({1\over a } + {1 \over a_-} \right) + \left({\dot{a}\over 2a }\right)^2 + \left({\dot{a} \over 2a_-}\right)^2 -  {1\over a }\del_+ \left({a_-\over a }\right) -  {1\over a_-}\del_- \left({a\over a_-}\right)   \right). 
\end{align}
We now note that the requirement $\del_-\dot a=0$ is equivalent to $a$ being of the form stated. Clearly, such a form obeys this condition as $\dot a=\alpha$ is constant on $\Z_n$. Conversely, given $a(t,i)$ obeying the condition, we let $\alpha(t)={1\over n}\sum_ia(t,i)$ be the average value and $\beta=a-\alpha$. The latter averages to zero and has zero time derivative by the assumption on $a$, hence depends only on $i$. We now insert this specific form  into the curvature calculations to obtain 
\begin{align}
	{\rm Ricci} =& \left( \frac{\ddot{\alpha}}{4} - \frac{s }{(\alpha+\beta)(\alpha+R_+\beta)}\right) e^+\tens e^-+\left( \frac{\ddot{\alpha}}{4} - R_-\left(  \frac{s }{(\alpha+\beta)(\alpha+R_-\beta)}\right) \right) e^-\tens e^+ \nonumber\\
	&-{\dot{\alpha}\over 4}R_-\left( \frac{\partial_+\beta}{(\alpha + \beta)^2}  \right) \extd t \tens e^+-  \frac{\partial_+\beta}{(\alpha + \beta)(\alpha + R_+\beta)}  e^+ \tens\extd t  \nonumber \\  
	&-{\dot{\alpha}\over 4}  \frac{\partial_-\beta }{(\alpha + \beta)^2} \extd t \tens e^-   
	-R_-\left(  \frac{\partial_-\beta}{(\alpha + \beta)(\alpha + R_-\beta)}  \right) e^- \tens\extd t  \nonumber\\
	&+\left( \frac{\ddot{\alpha}}{4} \left(\frac{2\alpha + \beta + R_-\beta}{(\alpha + \beta)(\alpha + R_-\beta)}\right) + \frac{\dot{\alpha}^2}{4}\left(  \frac{(\alpha+\beta+R_-\beta)^2-(\alpha^2+2\beta R_-\beta)}{(\alpha+\beta)^2(\alpha + R_-\beta)^2}   \right)   \right)\extd t \tens \extd t\end{align} and the scalar curvature as stated. Without loss of generality, we have fixed $\sum_i\beta(i)=0$ since this could be shifted into the value of $\alpha$. We also have  the geometric Laplacian 
	\begin{equation} \Delta f = -\Delta^{\Z_n} f - \left({1\over a} + {1\over a_-}\right){\dot{a} \over 2}\del_t f - \del_t^{2}f=- \left({1\over a} + {1\over a_-}\right)({\dot{a} \over 2}\del_t f -\Delta_{\Z_n}f)-\del_t^2f,\end{equation}
which simplifies as stated. We are using $\Delta^{\Z_n}$ for the Laplacian in Propostion~\ref{QLCZn} and $\Delta_{\Z_n}$ with lower label for the standard finite difference Laplacian. \endproof

In this theorem, $\alpha(t)>0$ is the average `radius' of the $\Z_n$ geometry, evolving with time, while $\beta(i)$ as a fluctuation as we go around $\Z_n$ and  we see that this has to be  `frozen' (does not depend on time) in order  for the metric to admit a quantum geometry. It is striking that this includes the FLRW-type models studied in the remaining section in the class forced by the quantum geometry. Note that we also need to restrict to 
\begin{equation}\label{constraint} {\rm min}_i\beta(i)>-{\rm inf}_t\alpha(t)\end{equation}
so that $a(t,i)$ is everywhere positive. 

Although we will not study it here, we are now in position to start thinking about quantum gravity on $\R\times\Z_n$ in a functional integral approach. Given the identified restrictions, this would presumably have the form of a partition function
\begin{equation}\label{ZFL} Z=\int{\mathcal D}\alpha \prod_{i=0}^{n-2}\int\!\!\! \extd\beta(i)\, J_\beta e^{{\imath \over \Gn }\int_{-\infty}^\infty\extd t\sum_{ \Z_n}\mu  S[\alpha,\beta] }\end{equation}
for some measure $\mu(t,i)$. Classically, the latter would come from the metric coefficients and, for example, we might take  something of the form $\mu=\sqrt{(\alpha+\beta)(\alpha+R_-\beta)}$ in line with the case of $\Z_n$ alone in Section~\ref{secqg}. It is not clear what would be the right choice, however. For the integral over functions $\{\alpha(t)\}$, there would be the usual issues to make this rigorous (as some kind of continuous product of integrals). The new feature is that these should be restricted to values $\alpha(t)>0$ and for a given configuration $\{\alpha(t)\}$, we should limit the lower bound on the $\int\extd \beta(i)$ integrations according to (\ref{constraint}). Finally, we presumably would want, to maintain the $\Z_n$ symmetry, a Jacobian which we have denoted $J_\beta$ to reflect the geometry of the constraint $\sum\beta(i)=0$.  The choice of $\mu$ and the constrained integration are both issues that we already encountered for $\Z_n$ in Section~\ref{secqg} but are now significantly more complicated.  We also should now aim for a physical theory given the Lorentzian signature, hence the $\imath$ in the action.

\subsection{Equations of state in FLRW model on $\R\times \Z_n$}\label{secFRW}

For the remainder of the paper, we focus on the cosmological FLRW model case where $a=\cp^2(t)$ with no fluctuation $\beta(i)$ over $\Z_n$ and hence
\begin{equation}\label{gFL} g=-\extd t\otimes\extd t- \cp^2(t) e^+\tens_s e^-,\end{equation}
where $e^+ \tens_s e^- = e^+ \tens e^- +e^- \tens e^+$.  In this case, the results above simplify to
\begin{align}
	\label{QLCFL}
	\nabla \extd t &= \cp\dot\cp e^+\tens_s e^-,\quad \nabla e^\pm=-  {\dot \cp\over \cp}e^\pm\tens_s\extd t,\\ 
	\label{RFL} 
	R_\nabla e^\pm&=-{\ddot \cp\over \cp}\extd t\wedge e^\pm\tens\extd t  \pm \left({\dot \cp\over \cp}\right)^2  \cp^2 e^+\wedge e^-\tens e^\pm,\quad 
	R_\nabla \extd t=  \ddot \cp \cp\extd t\wedge e^+\tens_s e^-,\\
	\label{RicFL} 
	{\rm Ricci}&= {\ddot \cp\over \cp} \extd t\tens \extd t+{1\over 2}\left({\dot\cp^2\over \cp^2}+{ \ddot\cp \over \cp}  \right) \cp^2 e^+\tens_s e^- ,\quad  S=-2{\ddot \cp\over \cp} -\left({\dot \cp\over \cp}\right)^2.
\end{align}

 Although a general scheme for a noncommutative Einstein tensor is not known, in the present model it seems sufficient to  define it in the usual way, in which case 
\begin{equation}
	\label{EinFL}
	{\rm Eins} = {\rm Ricci} - {1\over 2}Sg =  - {1\over 2 }\left( {\dot{\cp}\over \cp} \right)^2 \extd t\tens\extd t-{ \cp \ddot\cp  \over 2} e^+ \tens_s e^-. 
\end{equation}

\begin{lemma}\label{divergenceTwoForm}
	The divergence $\nabla\cdot=((\ ,\ )\tens\id)\nabla$ of a 1-1 tensor of the form
	\[ T=f \extd t\tens \extd t - p \cp^2 e^+ \tens_s e^-\] defined by functions $f,p$ on $\R\times \Z_n$, and for metric defined as above by $\cp(t)$,   is
	\begin{align*}
	 \nabla \cdot T = - \left(\dot{f} + 2{\dot{\cp}\over \cp} (f+p)\right)\extd t + \del_b p e^b. 
	\end{align*}
	In particular, the Einstein tensor (\ref{EinFL}) is conserved in the sense $\nabla\cdot {\rm Eins}=0$. 
\end{lemma}
\proof
The Leibniz rule for the action of the connection produces 
\begin{align} 
	\nabla(f \extd t&\tens \extd t - p \cp^2 e^+ \tens_s e^-)\nonumber \\
	&=  \extd f\tens \extd t \tens \extd t - \extd p\tens \cp^2 e^+ \tens_s e^- +f\nabla(\extd t \tens \extd t) - p\nabla(\cp^2 e^+ \tens_s e^-)\nonumber \\
	&=  \extd f\tens \extd t \tens \extd t - \extd p\tens \cp^2 e^+ \tens_s e^- + (f+p)\nabla(\extd t \tens \extd t)\nonumber \\ 
	&=  \dot{f} \extd t\tens \extd t \tens \extd t -  \dot{p} \extd t \tens \cp^2 e^+ \tens_s e^-  + \del_b f e^b \tens\extd t \tens \extd t +   \del_b p e^b\tens \cp^2 e^+ \tens_s e^-\nonumber\\
	&\quad +\cp\dot\cp(f+p)\left( e^+ \tens_s e^-\tens\extd t + e^-\tens \extd t \tens e^+ + e^+\tens \extd t \tens e^- \right)
\end{align}
on using metric compatibility whereby $\nabla(\extd t \tens \extd t) = -\nabla( \cp^2 e^+ \tens_s e^-)$ and then evaluating the former with $\sigma=$flip on $\extd t$. Now applying $(,)\tens \id$  with the inverse metric, we arrive at the stated result for the divergence. 

For ${\rm Eins}$ in (\ref{EinFL}), the coefficients are constant on $\Z_n$, so there is no $e^\pm$ term in $\nabla\cdot{\rm Eins}$. For the $\extd t$ term it is easy to verify that $\dot{f} + 2{\dot{\cp}\over \cp} (f+p)=0$ automatically for the effective values of the specific coefficients $f,p$ in (\ref{EinFL}) defined by $\cp(t)$.   \endproof

Next, recall from Section~\ref{seccurv} that our formulation of Ricci is -1/2 of the usual value, hence Einstein's equation for us should be written as 
\begin{equation}\label{eineq} {\rm Eins}+4\pi \Gn  T=0\end{equation}
 and from (\ref{EinFL}) we see that this holds if $T$ has the form for dust of pressure $p$ and densisty $f$, namely 
\begin{equation} T= p g + (f+p) \extd t\tens \extd t= f \extd t\tens \extd t- p \cp^2 e_+\tens_s e_-\end{equation}
for pressure and density
\begin{equation}
	\label{pfFL} 
	p=-{1\over  8 \pi \Gn }\left({\ddot \cp \over\cp}\right),\quad
	f={1\over  8 \pi \Gn }\left({\dot \cp\over \cp} \right)^2.
\end{equation}
Note that $T$ is automatically conserved by the same calculation as for the Einstein tensor and this does not give any constraint on $\cp(t)$. Setting \begin{equation}  \mathit{H}:= {\dot\cp\over \cp}, \end{equation} conservation is equivalent to the continuity equation \begin{equation} \dot{f} =- 2\mathit{H}(f+p), \end{equation}
which also holds automatically. The standard consideration in cosmology at this point is to assume an equation of state  $p = \omega f$ for a real parameter $\omega$, in which case the continuity equation becomes ${\extd f\over\extd \cp}=-2f(1+\omega)$ so that   $f \propto \cp^{-2(1+\omega)}$. Given this form of the density $f$, our assumption $p=\omega f$ can be solved for $\omega\ne -1$ to give
\begin{equation}
	\label{eq:FLRW_papram}
	 \cp(t) = \cp_0\left( 1 + \sqrt{8\pi G f_0}(1+w)t  \right)^{{1\over 1+w}}
\end{equation}
for initial radius and pressure $\cp_0,f_0$. Here $\omega>-1$ leads to an expanding universe. Recall that  one usually takes $\omega=0,1/3$ for cold dust and radiation respectively. 

If we add a cosmological constant so that ${\rm Eins} -{1\over 2}g\Lambda +4\pi \Gn  T=0$, this is
equivalent to a  modified stress energy tensor given as before but with modified
\begin{equation} f^\Lambda=f+{ \Lambda\over 8 \pi \Gn },\quad p^\Lambda=p-{ \Lambda\over 8 \pi \Gn }=\omega f^\Lambda- {1+\omega\over 8\pi \Gn }\Lambda. \end{equation}
The effective equation of state now leads to 
\begin{equation}\label{aFLLam} \cp(t)=\cp_0\left( \frac{\cosh({\rm arccosh} (\sqrt{-\frac{\Lambda}{8 \pi \Gn   f_0}})+\sqrt{\Lambda} (1+\omega) t)}{\sqrt{-\frac{\Lambda}{8 \pi \Gn   f_0}}}\right)^{1\over 1+\omega}\end{equation}
with reasonable behaviour for $f_0>0$ (with $f$ remaining positive) and real $\Lambda$ but a limited range of $t$ when $\Lambda<0$.   

For comparison, note that the classical Einstein tensor on $\R\times S^1$ with $g=-\extd t\tens\extd t+ \cp^2(t)\extd x\tens\extd x$ vanishes as for any 2-manifold and $T=f\extd t\tens\extd t+ p \cp^2(t) \extd x\tens \extd x=p g + (f+p)\extd t\tens\extd t$ admits only zero pressure and density if we want Einstein's equation. One can also add a cosmological constant, in which case we need $p=-{\Lambda\over 8\pi \Gn }$ and $f={\Lambda\over 8\pi \Gn }$ and $\omega=-1$. This is therefore not the right comparable. 

\begin{proposition} The results (\ref{eq:FLRW_papram})-(\ref{aFLLam}) for $\cp(t)$ (as well as for $f(t)$) for the FLRW  model on $\R\times\Z_n$ are the same as  for the classical flat FLRW-model on $\R\times \R^2$.
\end{proposition}
\proof The flat FLRW model in 1+2 dimensions is an easy exercise starting with the metric $g = -\extd t\tens\extd t + \cp^2(t) (  \extd x \tens \extd x + \extd y \tens \extd y)$ to compute the Ricci tensor  (in our conventions, which is $-{1\over 2}$ of the usual values) as
\begin{equation}{\rm Ricci} = { \ddot\cp \over \cp} \extd t	\tens \extd t - \frac{1}{2}\left( {\ddot\cp\over \cp} + {\dot\cp^2 \over \cp^2} \right)\cp^2(\extd x \tens \extd x + \extd y \tens \extd y)\end{equation}
and the same scalar curavture $S$ as in (\ref{RicFL}). The Einstein tensor is therefore 
\begin{equation} {\rm Eins}= - {1\over 2}\left( {\dot\cp\over \cp} \right)^2 \extd t\tens\extd t+{\cp\ddot\cp \over 2}(\extd x \tens \extd x + \extd y \tens \extd y)\end{equation}
by  a similar calculation as for (\ref{EinFL}). The stress tensor for dust being similarly $f\extd t\tens\extd t+ p \cp^2(\extd x \tens \extd x + \extd y \tens \extd y)$ means that the Einstein equations give $p,f$ by the same expressions (\ref{pfFL}) as before. The Friedmann equations are therefore  the same as we solved.  \endproof

This is perhaps not too surprising given that $\Omega^1$ on $\Z_n$ is 2-dimensional, indeed $-e^+\tens_s e^-$ plays the same role as the classical spatial metric $\extd x \tens \extd x + \extd y \tens \extd y$. We also recall  by way of comparison that the standard $k=0$ Friedmann equations for the FLRW model $\R\times \R^3$ has the well-known solution,
\begin{equation} \cp(t) = \cp_0(1 + \sqrt{6\pi \Gn  f_0}(w+1)t)^{\frac{2}{3(w+1)}}\end{equation}
without cosmological constant and can also be solved with it, as 
\begin{equation} \cp(t) = \cp_0 \left(\frac{\cosh{\left( {\rm arccosh\left( \sqrt{ -\frac{\Lambda}{8\pi \Gn  f_0}} \right)} + \sqrt{\frac{3\Lambda}{4}}(w+1)t \right)}}{\sqrt{-\frac{\Lambda}{8\pi \Gn  f_0}}}\right)^{\frac{2}{3(w+1)}}.\end{equation}

As usual, the case of $\cp(t)$ independent of time is a solution for the Einstein vacuum equation with ${\rm Ricci} = 0$. It is 
 easy to see that there are no other solutions of interest with
${\rm Ricci} \propto g$  or  ${\rm Eins} \propto g$. On the other hand, we do have the following. 

\begin{proposition}
The equation ${\rm Ricci}-\lambda S g=0$ with time-varying $\cp(t)$ and constant $\lambda$ has a unique solution of the form
	\[  \lambda={1\over 3},\quad  \cp(t)=\cp_0 e^{\mu t}\]
	for some growth constant $\mu \ne 0$ and initial  $\cp_0>0$. 
\end{proposition}
\proof 
Considering the equation ${\rm Ricci} = \lambda gS$, where $\lambda $ is an arbitrary real constant, we have two equations; one related to  $\twoForm{\pm}{\mp}$ is
\begin{equation}
	    \frac{\ddot\cp}{\cp}  + \left(\frac{2\lambda}{1-4\lambda} + 1\right)\left(\frac{\dot\cp}{\cp}\right)^2 = 0
\end{equation}
and other related to $\extd t \tens \extd t$ is
\begin{equation}
	  \frac{\ddot\cp}{\cp} + \left( \frac{\lambda-1}{1-2\lambda} + 1\right)\left(\frac{\dot\cp}{\cp}\right)^2 = 0.
\end{equation}
This requires $\lambda = {1\over 3}$ and  $\frac{\ddot\cp}{\cp} = \left(\frac{\dot\cp}{\cp}\right)^2$,  which has the solution claimed.
\endproof

\subsection{Quantum field theory on $\R\times\Z_n$}\label{secqftR} Here we consider quantum field theory in the flat case where $\cp$ is a constant. The corresponding Laplacian operator and the Klein-Gordon equation are
 \begin{equation} \Delta = {2\over \cp^2} \left( \partial_+ + \partial_- \right) - \partial_t^2 ; \quad (-\Delta + m^2)\phi=0.\end{equation}
 We write $q=e^{2\pi\imath\over n}$, where $\imath$ denotes the imaginary unit, and Fourier transform on $\Z_n$ by considering solutions of the form $\phi(t,i)=q^{ik}e^{-\imath w_k t}$, where $i$ denotes the position in $\Z_n$. This is labelled by a discrete momentum $k=0,\cdots,n-1$ with associated `mass on-shell' expression 
 \begin{equation}
	\label{mass-onshell}
	w_k^2 = {8\over \cp^2} \sin^2{\left({\pi\over n} k\right)} +m^2.
 \end{equation}
We then consider the corresponding operator-valued fields starting with
\begin{equation} \phi_i= \sum_{k = 0}^{n-1} {1\over \sqrt{2 w_k}} \(q^{ik}a_k + q^{-ik}a_k^{\dag}\), \end{equation}
where now $a_k, a^\dag_k$ are self-adjoint operators and $a_k\ket{0} = 0$, with $\ket{k}$ eigenvectors of the corresponding Hamiltonian
\begin{equation} H = \sum_{k = 0}^{n-1} w_k( a_ka_k^{\dag} +{n\over 2}). \end{equation}
From the commutators $[H,a_k]=-w_k a_k$ and $[H,a_k^\dag]=w_k a_k^\dag $, and using the Heisenberg representation for the time evolution of the field, we obtain
\begin{equation}\label{qftphi} \phi_i(t) = e^{\imath Ht}\phi_i e^{- \imath Ht} = \sum_{k = 0}^{n-1} {1\over \sqrt{2 w_k}} \( q^{ik-\imath w_kt}a_k + q^{-ik+\imath w_kt}a_k^{\dag} \) \end{equation} with the time-ordered correlation function
\begin{equation}
 \bra{0} T[\phi_i(t_a) \phi_j(t_b)]\ket{0} = \sum_{k = 0}^{n-1}{1\over w_k} \cos{\left({2\pi \over n}k(i-j)\right)}e^{-\imath w_k |t_a - t_b|}. 
\end{equation}

Next we check that we obtain the same  correlation function via a formal path integral approach with the $\imath\epsilon$-prescription.  The  partition functional integral $Z[J]$ with source $J$ is defined as
\begin{equation} Z[J] = {\int {\mathcal D}\phi\, e^{ {1\over \beta}S[\phi] + {1\over \beta}\int \sum_{i=0}^{n-1} J_i(t)\phi_i(t) }   \over \int {\mathcal D}\phi\, e^{ {1\over \beta}S[\phi] } }  =  {\int {\mathcal D}\phi\, e^{ {1\over 2\beta} \int dt \sum_{i = 0}^{n-1} \left( \phi_i(t)(\Delta - m^2 + \imath\epsilon)\phi_i(t) + 2J_i(t)\phi_i(t) \right)}\over \int {\mathcal D}\phi\, e^{ {1\over 2\beta} \int dt \sum_{i = 0}^{n-1} \left( \phi_i(t)(\Delta - m^2 + \imath\epsilon)\phi_i(t) \right)} }, \end{equation} where $\beta$ is a dimensionless coupling constant. We diagonalize the action $S[\phi]$ using Fourier transform to write \begin{equation}  \phi_i(t) = \sum_{k = 0}^{n-1}\int_{-\infty}^\infty {dw\over 2\pi}  \tilde{\phi}_k(w) q^{ik} e^{\imath wt}; \quad J_i(t) = \sum_{k = 0}^{n-1}\int_{-\infty}^\infty {dw\over 2\pi}  \tilde{J}_k(w) q^{ik} e^{\imath wt}, \end{equation}
which produces the action
\begin{equation} S[\tilde{\phi}] = \int_{-\infty}^{\infty} {dw\over 2\pi}{1\over 2\beta} \sum_{k=0}^{n-1} \left(   \tilde{\phi'}_{-k}(-w) (-w^2 + w_k^2)\tilde{\phi'}_k(w) + \tilde{J}_{-k}(-w)  {1\over -w^2 + w_k^2} \tilde{J}_k(w)\right),    \end{equation}
where $\tilde{\phi'}_k(w) = \tilde{\phi}_k(w) - (-w^2 + w^2_k)^{-1}\tilde{J}_k(w)$. The first term in terms of the new variables gives a Gaussian integral, which we ignore as an overall factor independent of the source. Using  
\begin{equation} \tilde{J}_k(w) ={1\over n} \int dt \sum_{i=0}^{n-1}  J_i(t) q^{-ik}e^{\imath wt}, \end{equation}
 the functional integral becomes 
\begin{equation}Z[J] = e^{{1\over \beta}\int  dt' dt'' J_i(t') \imath\Delta_f(i,t';j,t'')J_j(t'')}, \end{equation}
where the Feynman propagator is 
\begin{align}
	\Delta_f(i,t';j,t'') &= \sum_{k = 0}^{n-1}q^{k(i-j)}  \int {dw\over 2\pi} {e^{-\imath w(t'-t'')} \over (-w+w_k-\imath\epsilon\)(w + w_k+ \imath\epsilon)}\nonumber \\
	&= \sum_{k = 0}^{n-1}{1\over w_k} \cos{\left({2\pi \over n}k(i-j)\right)}e^{-\imath w_k |t_a - t_b|}.
\end{align}
Finally, by construction, we have
\begin{equation}  \bra{0} T[\phi_i(t_a) \phi_j(t_b)]\ket{0} = {\beta^2\over \imath^2} {\partial\over\partial J_i(t_a)} {\partial\over\partial J_j(t_b)}Z[J]=\Delta_f(i,t';j,t''), \end{equation} 
which therefore gives the same result as obtained by Hamiltonian quantisation. This is as expected, but provides a useful check that our methodology makes sense at least in the flat case of constant $\cp$.

\subsection{Particle creation in FLRW model on $\R\times\Z_n$}\label{sechawR}

Here we follow the procedure developed by Parker \cite{Par69, Par12, ParNav,ParTom} to study cosmological particle creation, adapted now to an FLRW model on $\R\times\Z_n$ with an expanding quantum metric (\ref{gFL}). 

\subsubsection{Model case of $\R\times S^1$.} We start with the classical background geometry case of $\R\times S^1$, which is presumably known but sets up the procedure and our notations. Here the metric has the usual 2D FLRW form
\begin{equation} g = -\extd t\tens \extd t + \cp^2(t) \extd x\tens\extd x, \end{equation}
where $\cp(t)$ is an arbitrary positive function. Thus the Klein-Gordon equation for the field $\phi$ is
\begin{equation} \left(g^{\mu\nu}\nabla_\mu\nabla_\nu -m^2\right)\phi = 0 \end{equation}
or in explicit form
\begin{equation}
\ddot{\phi} + \frac{\dot\cp}{\cp}\dot{\phi} - \frac{1}{\cp^2}\partial^2_x\phi + m^2\phi = 0. 
\label{eq:k-g-class}
\end{equation}
We impose the periodic boundary condition  $\phi(t,x + L) = \phi(t,x)$, where $L$ is a dimensionless parameter for the normalisation of the box geometry. We then expand the field in terms of a Fourier series

\begin{equation}
\label{eq:first-exp}
\phi(t,x) = \sum_k (A_k f_k(t,x) + A_k^* f^*_k(t,x) ), 
\end{equation}
where
\begin{equation}
	f_k(t,x) = \frac{1}{\sqrt{L\cp}}e^{\imath xk}h_k(t) 
	\label{eq:f_k}
\end{equation}
and $k = 2l\pi/L$ for $l$ an integer. Here $k/R$ is the physical momentum and $l$ the corresponding `integer momentum' on a circle. Then $\phi$ obeys (\ref{eq:k-g-class}) provided 
\begin{equation}
\label{eq:h_k}
\ddot{h}_k(t) + \left(   \frac{k^2}{\cp^2} +  m^2   \right)h_k(t) + \left(\frac{1}{4}\left( \frac{\dot\cp}{\cp} \right)^2 - \frac{1}{2}\frac{\ddot\cp}{\cp} \right)h_k(t) = 0
\end{equation}
for each momentum mode. We will be particularly interested in the adiabatic limit, where $\cp$ varies slowly with respect to the time in such way that $\dot\cp/\cp \rightarrow 0, \ddot\cp / \cp \rightarrow 0$. The solutions to (\ref{eq:h_k}) in this approximation are
\begin{equation}
\label{eq:liu-aprox}
h_k(t) \sim (w_k) ^{-\frac{1}{2}} \left(\alpha_k e^{\imath \int^t  w_k(t') \extd t'}  + \beta_ke^{-\imath \int^t  w_k(t') \extd t'} \right), 
\end{equation}
where $\alpha_k$ and $\beta_k$ are complex constants that satisfy
\begin{equation}
\label{con:bog}
|\alpha_k|^2 - |\beta_k|^2 = 1
\end{equation}
and
\begin{equation}
\label{eq:w_k}
w_k(t) = \sqrt{ m^2+\frac{k^2}{\cp^2(t)} }. 
\end{equation}

In order to have an exact solution, we now let $\alpha_k$ and $\beta_k$ be functions of time such that
\begin{equation}
\label{eq:liu-exact}
h_k(t) = (w_k(t)) ^{-\frac{1}{2}} \left(\alpha_k(t) e^{\imath \int^t  w_k(t') \extd t'}  + \beta_k(t)e^{-\imath \int^t  w_k(t') \extd t'} \right)
\end{equation}
and 
\begin{equation}
\label{condition_bolug}
|\alpha_k(t)|^2 - |\beta_k(t)|^2= 1
\end{equation}
for all $t$. Equivalently,  we can rewrite the expansion of the field as
\begin{equation}
\label{eq:sec-exp}
\phi(t,x) = \sum_k (a_k(t) g_k(t,x) + a_k^*(t) g^*_k(t,x) ), 
\end{equation}
where now
\begin{equation} g_k(t,x)  =  \frac{\cp^{-\frac{1}{2}}}{\sqrt{Lw_k}} e^{\imath (xk - \int^t w_k(t')\extd t' )} \end{equation}
and
\begin{equation}
\label{eq:a_operator}
a_k(t) = \alpha_k(t) ^*A_k + \beta_k (t) A^*_k.  
\end{equation}

In order to follow the usual procedure of canonical quantisation, we next define the conjugate momentum as
\begin{equation} \pi(t,x) = \cp \dot{\phi}(t,x), \end{equation}
promote the field $\phi(t,x)$ and the momentum $\pi(t,x)$ to operators $\hat{\phi}(t,x), \hat{\pi}(t,x)$ respectively, and  impose the commutators relations
\begin{equation}
\label{eq:can_com}
[\hat\phi(t,x),\hat\phi(t,x')] =  [\hat\pi(t,x),\hat\pi(t,x')] = 0, \quad [\hat\phi(t,x), \hat\pi(t,x')] = \imath\delta(x-x'). 
\end{equation}
This requires that  $A_k$ and $A_k^*$ in (\ref{eq:a_operator}) are promoted to operators $A_k$ and $A_k^\dagger$ with the usual commutation relations
\begin{equation}
[ A_{k'}, A_k ] = [A_{k´}^\dagger,A_{k'}^\dagger] = 0, \quad [A_{k'},A_k^\dagger] = \delta_{k,k'}.
\end{equation}
It then follows from these and a  conserved quantity (see \cite{Par69}),  that the operator versions of  (\ref{eq:a_operator}) obey 
\begin{equation}
\label{eq:gen_com}
[a_k(t), a_{k'}(t)] = [a^\dagger_k(t), a^\dagger_{k'}(t)] = 0, \quad   [a_k(t), a^\dagger_{k'}(t)] = \delta_{k,k'}. 
\end{equation}

Now note that for any function $W_k(t)$ with at least derivatives to second order, the function
\begin{equation}
\label{eq:general_aprox}
H(t):= W_k(t){}^{-{1\over 2}}(\alpha_ke^{\imath \int^t dt' W_k(t')}  + \beta_k e^{-\imath \int^t dt' W_k(t')}) 
\end{equation}
for any constants $\alpha_k,\beta_k$ is an exact solution of the equation
\begin{equation} \ddot{H}(t) + \left[  W_k^2 - W_k^{\frac{1}{2}}\frac{d^2}{dt^2}W_k^{-\frac{1}{2}} \right]H(t) = 0. \end{equation}
Hence, if we can solve for $W_k(t)$ such that 
\begin{equation}\label{eq:W_second_order}
W_k^2 =   W_k^{\frac{1}{2}}\frac{d^2}{dt^2}W_k^{-\frac{1}{2}} + w_k^2  + \sigma
\end{equation}
holds, where 
\begin{equation} \sigma = \frac{1}{4}\left( \frac{\dot\cp}{\cp} \right)^2 - \frac{1}{2}\frac{\ddot\cp}{\cp}, \end{equation}
then $H(t)$ provides exact solutions $h_k(t)$ of  (\ref{eq:h_k}) for each $k$. 

We can then expand $W_k$ as a sum of terms
\begin{equation}
\label{eq:W_expantion}
W_k = w^{(0)} + w^{(1)}+ w^{(2)} +\dots,  
\end{equation} 
where the superfix denotes the adiabatic order. Putting this into (\ref{eq:W_second_order}) and just keeping the elements of order zero, we have $w^{(0)}= w_k$. Just keeping the elements of first order tell us that $w^{(1)}=0$, while for elements of second adiabatic order we require
\begin{equation}  w^{(2)} =  \frac{(w^{(0)})^{-\frac{1}{2}}}{2} \frac{d^2}{dt^2}\left((w^{(0)})^{- \frac{1}{2}}\right)  + \frac{\sigma}{2w^{(0)}}.\end{equation}
We can continue this procedure to any desired order to find odd $w^{(i)}=0$ and even $w^{(i)}$ determined from lower even ones.  The form of the functions $\alpha_k(t)$ and $\beta_k(t)$ can be obtained when we impose  (\ref{condition_bolug}). From its temporal derivative, one is led to the ansatz 
\begin{equation}\label{alphaW}
	\alpha_k(t) = - \dot{\beta_{k}} (t)e^{-2\imath\int^t dt' W_k(t')},\quad \beta_k(t) =  -\dot{\alpha_{k}} (t)e^{2\imath\int^t dt' W_k(t')} 
\end{equation}
as justified by consistency with (\ref{eq:h_k}),  given (\ref{eq:W_second_order}). For a more explicit form of these coefficients,  see \cite{Par65}.

A special case of interest here is when the $w_k^{(i)} $ vanish for all the orders bigger that zero (and all  $k$). In this case,   the operator $a_k(t)$ defined in (\ref{eq:a_operator}) is independent of time,  the number of particles is constant and there is no particle creation. From the above remarks, it is sufficient that $w_k^{(2)}=0$, which amounts to
\begin{equation}\label{nopart-FL}
\frac{1}{4}\frac{m^2\left(4\frac{k^2}{\cp^2} - m^2 \right)}{\left(  \frac{k^2}{\cp^2} + m^2 \right)^2}\left(\frac{\dot\cp}{\cp}\right)^2 + \frac{1}{2} \frac{m^2}{(\frac{k^2}{\cp^2} + m^2) } \frac{\ddot\cp}{\cp} = 0. 
\end{equation}
The only way that this can hold for all time and $k$ is in the infinite mass limit $m\rightarrow \infty$ (cf. \cite{Par69}), where it reduces to an FLRW-like equation
\begin{equation} \frac{1}{2}\frac{\ddot\cp}{\cp}= \frac{1}{4}\left(\frac{\dot\cp}{\cp}\right)^2\end{equation}
with solution  $\cp\propto t^{2}$. As well as the obvious flat Minkowski case of constant $\cp$, this represents a further possibility for  no particle creation. 

For an actual particle creation computation, it is convenient to move to a new time variable $\eta$ such that 
\begin{equation}\label{etat}
\extd\eta =\frac{ \extd t}{\cp(t)},
\end{equation}
in which case our metric becomes conformally flat as
\begin{equation}
	g = C(\eta)(- \extd \eta \tens \extd \eta + \extd x \tens \extd x),
\end{equation}
where $C(\eta) = \cp^2(t)$ is now regarded as a function of $\eta$. Following the same steps as before but using this metric puts the wave equation  (\ref{eq:h_k}) on spatial momentum modes in the simpler form 
\begin{equation}
\label{eq:h_k_flat}
\frac{\extd^2h_k(\eta) }{\extd\eta^2} + w_k(\eta) h_k(\eta) = 0,
\end{equation}
where
\begin{equation}
	w_k(\eta) = \sqrt{C(\eta) m^2 + k^2}
	\label{eq:freq_circle}
\end{equation}
as a modification of (\ref{eq:w_k}). 

We now consider particle creation under the assumption that $\cp$ and hence $C$ has a constant constant value $C(\eta)=\cp^2_{in}$ for early times $\eta<\eta_{in}$, say, and a constant value  $C(\eta)=\cp^2_{out}$ for late times $\eta>\eta_{out}$, with  $\eta_{in} < \eta_{out}$. For these early and late times, we let 
\begin{equation}\label{circlewkin}
	w_k^{\rm in}= \sqrt{\cp^2_{in} m^2+k^2 } ; \quad w_k^{\rm out}= \sqrt{\cp^2_{out} m^2+k^2 } 
\end{equation}
as functions of $k$. The fields at early and late times behave exactly as flat Minkowski space-time with the corresponding frequency or effective mass, with solutions of (\ref{eq:h_k_flat}) at early and late times provided by
\begin{equation}
 h_k^{\rm in}(\eta)= (w_k^{\rm in})^{-\frac{1}{2}} e^{\imath w_k^{\rm in} \eta},\quad 
	h_k^{\rm out} (\eta)=(w_k^{\rm out})^{-\frac{1}{2}} e^{\imath w_k^{\rm out} \eta}. 
\end{equation}
Now suppose that we start with $h_k^{\rm in}(\eta)$ at early times, i.e. $h_k(\eta)$ for $\alpha_k(\eta_{in})=1$ and $\beta_k(\eta_{in})=0$ in the analogue of (\ref{eq:liu-exact}), and extend this by solving (\ref{eq:h_k_flat}) to late times. There we expand it as the Bogolyubov transformation 
\begin{equation}\label{eq:bog} h_k^{\rm in}= \alpha_{k} h_k^{\rm out} + \beta_{k} h_k^{\rm out}{}^*\end{equation} 
valid at late times and for some complex constants $\alpha_k$, $\beta_k$. Comparing with the analogue of (\ref{eq:liu-exact}) at late times, these constants up to phases are just the evolved values $\alpha_k(\eta_{out}),\beta_k(\eta_{out})$ in the general scheme. (The  phases come from $e^{\imath\int_{\eta_{in}}^{\eta_{out}}w_k(\eta)\extd\eta}$ and are not relevant in what follows.) 

Finally, we fix a vacuum $|0\>$ as characterised by  $A_k\ket{0} = 0$ and consider the number operator $N_k(\eta)=a_k^\dagger( \eta) a_k( \eta)$ is it evolves in time, where we use the analogue of  (\ref{eq:a_operator}) as our solution evolves. Starting now with  $\alpha_k( \eta_{in})=1$, $\beta_k( \eta_{in})=0$ in defining $a_k,a_k^\dagger$, we have of course 
\begin{equation} \bra{0}N_k(\eta_{in})\ket{0}=0\end{equation}
at early times, but in this same state at late times we have the possibility of particle creation according to
\begin{equation}
\<N_k\>:=\bra{0}N_k(\eta_{out})\ket{0} = |\beta_{k}(\eta_{out})|^2=|\beta_k|^2.
\end{equation}

This completes the general scheme, which is also well-known from several other points of view. To proceed further we need to fix a particular $C(\eta)$, and the standard choice for purposes of calculation is to interpolate the initial and final values as
\begin{equation}
\label{eq:C}
C(\eta) =  \frac{\cp^2_{in}+\cp^2_{out}}{2} + \frac{\cp^2_{out}-\cp^2_{in}}{2}\tanh(\mu\eta), 
\end{equation}
where $\mu$ is a positive constant parameter.   Equation (\ref{eq:h_k_flat}) can then be solved with hypergeometric functions that have the correct asymptotic limit for late and early times. Comparison with (\ref{eq:bog})  gives (see \cite{Birrel}), 
\begin{gather}
	\alpha_{k} = \left(\frac{w_{k}^{\rm out}}{w_{k}^{\rm in}}\right)^{1/2}\frac{\Gamma(1-\imath\frac{ w_{k}^{\rm in}}{\mu})  \Gamma(-\imath\frac{ w_{k}^{\rm out}}{\mu} )}{\Gamma(-\imath\frac{ w_{k}^+}{\mu})  \Gamma(1-\imath\frac{ w_{k}^+}{\mu} )},\\  
\beta_{k} =  \left(\frac{w_{k}^{\rm out}}{w_{k}^{\rm in}}\right)^{1/2}\frac{\Gamma(1-\imath\frac{ w_{k}^{\rm in}}{\mu})  \Gamma(\imath\frac{ w_{k}^{\rm out}}{\mu} )}{\Gamma(\imath\frac{ w_{k}^-}{\mu})  \Gamma(1+\imath\frac{ w_{k}^-}{\mu} )}, \end{gather}
where
\begin{align}
	w_k^{\pm} = \frac{1}{2}(w_k^{out}\pm w_k^{in}). 
\end{align}
These values result in
\begin{align}
	|\alpha_k|^2 = \frac{\sinh^2{\left(\pi \frac{w_k^+}{\mu}\right)} }{\sinh{(\pi \frac{w_k^{in}}{\mu}  )} \sinh{(\pi \frac{w_k^{out}}{\mu})}},\quad |\beta_k|^2 = \frac{\sinh^2{\left(\pi \frac{w_k^-}{\mu}\right)} }{\sinh{(\pi \frac{w_k^{in}}{\mu}  )} \sinh{(\pi \frac{w_k^{out}}{\mu})}},  \label{eq:beta}
\end{align}
which, as one can check, obeys the unitarity condition (\ref{condition_bolug}). Figure \ref{fig:pol-circle} includes a plot of   $\braket{N_k} = |\beta_k|^2$ as a function of $k$, or rather of the associated integer momentum $l$. 

\subsubsection{Adaptation to $\R\times\Z_n$.} 

We now repeat the previous analysis for the polygon case with $n$ sides and time-varying metric (\ref{gFL}). We have the Laplacian 
\begin{equation}\label{LapFL}  \Delta = - \del^2_t -2\frac{\dot\cp}{\cp}\del_t+\frac{2}{\cp^2}(\partial_{+} + \partial_{-})\end{equation}
from Theorem~\ref{thm:connection} with $\beta=0$. The Klein-Gordon equation $(-\Delta + m^2)\phi = 0$ is 
\begin{equation}
\label{eq:k-g-discrete}
\left(-\frac{2}{\cp^2} (\partial_+ + \partial_-) + \frac{1}{\cp^2}\partial_t(\cp^2\partial_t) +m^2\right)\phi = 0. 
\end{equation}

Next, we expand the field in terms of a Fourier series
\begin{equation}
\label{eq:first-exp-Zn}
\phi(t,i) = \sum_k (A_k f_k(t,i) + A_k^* f^*_k(t,i) ) 
\end{equation}
in place of (\ref{eq:first-exp}), where now
\begin{equation}
f_k(t,i) = \frac{1}{\cp(t)}q^{ik}h_k(t) 
\end{equation}
and $k$ is an integer mod $n$. For the modes $f_k$  to obey (\ref{eq:k-g-discrete}), the $h_k$  have to solve
\begin{equation}
\ddot{h}_k(t) + \left( m^2 + \frac{8}{\cp^2}\sin^2{\left(\frac{\pi}{n}k\right)  }  \right)h_k(t)  - \frac{\ddot\cp}{\cp} h_k(t) = 0. 
\label{fao:eq}
\end{equation}
The corresponding on-shell frequency is therefore 
\begin{equation}
\label{eq:w_k_dis}
w_k(t) = \sqrt{  m^2 + \frac{8}{\cp^2(t)}\sin^2{\left(\frac{\pi}{n}k\right)  }  }
\end{equation}
instead of (\ref{eq:w_k}). We again consider an exact solution of the form
\begin{equation}
\label{eq:liu-exact-dis}
h_k(t) = (w_k(t)) ^{-\frac{1}{2}} \left(\alpha_k(t) e^{\imath \int^t  w_k(t') \extd t'}  + \beta_k(t)e^{-\imath \int^t  w_k(t') \extd t'} \right).
\end{equation}

Analogously to the previous case, we can re-write the expansion of the field as
\begin{equation}
\label{eq:sec-exp-Zn}
\phi(t,i) = \sum_k (a_k(t) g_k(t,i) + a_k^*(t) g^*_k(t,i) ), 
\end{equation}
where
\begin{equation} g_k(t,i)  =  {\cp^{-1}\over\sqrt w_k}  q^{ik} e^{-\imath\int^t w_k(t')\extd t' } \end{equation}
and the operator $a_k(t)$ has the same form as (\ref{eq:a_operator}). The quantisation procedure and analysis then proceeds as before. Our previous expressions for $W_k(t),\alpha_k(t), \alpha_k$ are still valid, but we have to take into account that the zero  adiabatic order term $w_k$ is different and that now
\begin{equation} \sigma=  -\frac{\ddot\cp}{\cp} \end{equation}
as the factor in (\ref{fao:eq}). 

For our first result, we look at when the $w_k^{(2)}$ correction vanishes so that there is no particle creation. In place of (\ref{nopart-FL}), we now require
\begin{equation}
\label{eq:w_2}
\frac{  \frac{4}{\cp^2}\sin^2{\left(  \frac{\pi}{n}k\right)}(\frac{4}{\cp^2}\sin^2{\left(  \frac{\pi}{n}k\right)} +  3m^2)  }{\left( \frac{8}{\cp^2}\sin^2{\left(\frac{\pi}{n}k\right)} + m^2  \right)^2  } \left(\frac{\dot\cp}{\cp}\right)^2  +  \frac{  \frac{4}{\cp^2} \sin^2{\left(\frac{\pi}{n}k\right)}  +m^2}{\left(\frac{8}{\cp^2}\sin^2{\left(\frac{\pi}{n}k\right)} + m^2\right) }\frac{\ddot{\cp}}{\cp} = 0.
\end{equation}
This can happen for all time and all $k$ in the infinite mass limit $m\rightarrow \infty$ if
\begin{equation}    \ddot\cp = 0 \end{equation}
with solution  $\cp\propto t$. However, we {\em also} have a new possibility when $m\to 0$, with 
\begin{equation}    \frac{\ddot\cp}{\cp} = -\frac{1}{2}\left(\frac{\dot\cp}{\cp}\right)^2\end{equation}
and solution  $\cp\propto t^{\frac{2}{3}}$. Thus we have not one but two additional possibilities for no particle creation beyond the constant Minkowski metric case. 

For our second result, we want to analyse particle creation for the $\R\times\Z_n$ model in an analogous way to the case when space is a circle. Thus, we make the same change of variable (\ref{etat}) in the metric (\ref{gFL}) to write
\begin{equation}
	g = C(\eta)(-\extd\eta\tens \extd\eta - e^+\tens_s e^-), 
\end{equation}
where $C(\eta) = \cp^2(t)$, and the corresponding connection is
\begin{align}
	\nabla \extd \eta = \frac{\dot\cp}{\cp} (-\extd \eta \tens \extd \eta + e^+\tens_s e^-), \quad \nabla e^{\pm} = -\frac{\dot\cp}{\cp}e^\pm\tens_s \extd \eta. 
\end{align}

\begin{figure}
	\includegraphics[scale=.75]{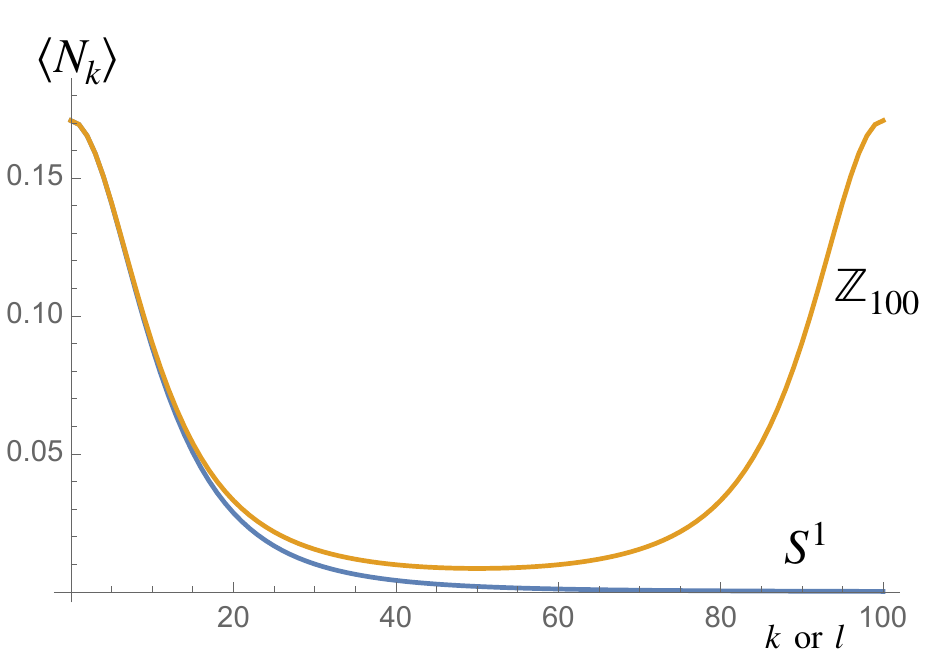}
	\caption{Number operator for $\Z_{100}$ against $k$ compared to $S^1$ with length scale factor $L=100/\sqrt{2}$, plotted against integer momentum $l$ where $k=2\pi l/L$. In both cases,  $R_{\rm in}\, m= 1$, $R_{\rm out}\, m=\sqrt{5}$ and $\mu=100$ for the interpolation parameter.}
	\label{fig:pol-circle}
\end{figure}

Using the quantum geometric Laplacian for this connection, we require
\begin{equation}
	\label{eq:h_k_flat_pol}
	\frac{\extd^2h_k(\eta) }{\extd\eta^2} + \left(C(\eta) m^2 + 8\sin^2{\left(\frac{\pi}{n}k\right)  } \right)h_k(\eta) = 0
\end{equation}
analogously  to (\ref{eq:h_k_flat}), but now in place (\ref{eq:freq_circle}) we have
\begin{equation}
	\label{eq:freq_pol}
	w_k(\eta) = \sqrt{ C(\eta) m^2 + 8\sin^2{\left(\frac{\pi}{n}k\right)  }  }. 
\end{equation}
The rest of the procedure follows in the same way with the same considerations, and in particular (\ref{eq:beta}) is still valid but with (\ref{eq:freq_pol}) instead of (\ref{eq:freq_circle}). Figure~\ref{fig:pol-circle} shows the expected value of the number operator $\braket{ N_k}$ as a function of $k$ as well as comparing to the circle case. The big difference of course is that the $\Z_n$ has to be periodic in $k$ since this is only defined mod $n$. 

\section{Concluding Remarks}\label{secRem}

In Section~\ref{secpoly}, we completely solved the quantum Riemannian geometry on a polygon $\Z_n$ in the sense of arbitrary square-lengths $a(i)$ on the edges. As is typical for discrete calculi, the increasing and decreasing derivatives are closely related but nevertheless linearly independent so that $\Omega^1$ is 2-dimensional -- in effect, the polygon acquires an extra `normal' direction (a remnant of a quantum geometry effect) and now admits curvature. Clearly, one could look beyond to discrete tori $\Z_{n_1}\times\cdots\times \Z_{n_m}$ and as well as to electromagnetism both in flat and curved metrics on the $\Z_{n_i}$ factors. Also interesting could be quantum geodesics even on one copy $\Z_n$, using the new formalism of \cite{BegMa:geo}. 

We also exploited the functorial nature of the formalism to take the continuum limit of the discrete geometry on $\Z_n$ in Section~\ref{seccirc}, first converting to a $q$-deformed geometry on the reduced circle $\C[s,s^{-1}]$ with $s^n=1$ and $q^n=1$, and then dropping the restriction on $s$ while sending $q\to 1$. We arrived in Corollary~\ref{circlim} at a central extension by a 1-form $\Theta_0$ of the classical differential forms on an algebraic circle, which can then be embedded in a $C^\infty(S^1)$ version with $s=e^{\imath\theta}$  using a formalism in \cite[Chaps.~1.3, 8.3]{BegMa:book}\cite{Ma:alm,Ma:rec}. We demonstrated how the continuum metric could also emerge, focussing on the constant $a=1$ case to illustrate the remaining issues. Specifically, the discrete metric had to be rescaled and expanded at the $q$-deformed level as
\begin{equation} g_0=(q-q^{-1})^2(e^+\tens e^-+e^-\tens e^+)= - 2 f^+\tens f^++ O(\Theta_0)\end{equation}
where $f^+=s^{-1}\extd s$ projects by setting $\Theta_0\to 0$ onto the 1-dimensional classical circle, so the first term projects to $2\extd\theta\tens\extd\theta$. The scale factor $(q-q^{-1})^2= -4 \sin^2({2\pi\over n})$ in the $\Z_n$ case is negative, which explains why, counterintuitively from the graph point of view, the physical metric needed an overall minus sign in later sections. However, some of the coefficients in the $O(\Theta_0)$ terms are singular as $q\to 1$ and we had to assume that they remain killed by $\Theta_0\to 0$. To resolve this would need some significant functional analysis in order to formulate the limiting process more carefully, which was beyond our scope here.  It would also be interesting to extend these ideas to more complicated models where a family of discrete approximations of a Riemannian manifold $M$ may limit to a one-higher dimension central extension of the classical geometry of $C^\infty(M)$ of the type in \cite{Ma:alm} (where the central extension formalism was used as a wave-operator approach to a noncommutative black hole). The discrete quantum geometry in our approach works {\em in principle} for any graph\cite{Ma:gra}, not only Cayley graphs on a discrete group, but while there is always a `maximal prolongation'\cite[Lem.~1.32]{BegMa:book} candidate for $\Omega^2$ and higher forms, for a reasonable continuum limit we will need to cut this down according to the manifold that we are approximating and so as to be able to solve for a quantum Levi-Civita connection. A first step would be to construct quantum geometries for general metrics on some other interesting  graphs beyond the group case, which remains substantially open. 

We then, in Section~\ref{secqg}, computed Euclideanised quantum gravity expectation values on $\Z_n$ for small $n$. In the spirit of $\Z_2\times\Z_2$ in \cite{Ma:sq}, we did this in two versions: the full quantisation and one for only fluctuations relative to an average field value. The polygon case is very different in that the full quantisation in terms of the ratios $\rho_i=a(i+1)/a(i)$ that enter into the action appears to be finite, but numerical work for $n=3$ gave us a strikingly similar phenomenon for the uncertainty $\Delta\rho_i\sim 1.1\<\rho_i\>$ (compared to $\Delta a=\<a\>/\sqrt{8}$ in  \cite{Ma:sq}). It was speculated in \cite{Ma:sq} that this could be indicative of some kind of vacuum energy. The metric correlation functions on $\Z_n$ were also substantial enough now to be interesting. These were computed more fully in the relative theory, where we found it useful to work with $b_i=a(i)/A$, with $A$ the {\em geometric mean}  of the $a$ field values rather than the additive one as in \cite{Ma:sq}. These results, in Figure~\ref{figZn}, are somewhat similar to correlations for a scalar field lattice box in \cite{Ma:haw}, but now in a real positive version, which both reassures us that the model is giving reasonable answers and gives a flavour of what to expect for quantum gravity in our approach. Clearly, more baby models should be computed to develop our intuition further. As discussed in \cite{Ma:sq}, our approach is not immediately comparable with other computable approaches such as \cite{Loll,Ash,Dow,Hale}. 

We also looked in Section~\ref{seccross} at the quantum geometry on $\R\times\Z_n$, including a first look at quantum gravity now with a time direction $\R$. The most striking result is that centrality of the quantum metric forces the shift vectors to vanish so that the quantum metric is block diagonal with the metric on $\Z_n$ free as before but scaled to an average value which can depend on time, see Theorem~\ref{thm:connection}. This is another example of the phenomenon in \cite{BegMa:gra} that not every classical metric can be the limit of a quantum geometry due to the centrality constraint. The general phenomenon here appears at the Poisson level \cite[Chap.~9.6]{BegMa:book}\cite{BegMa:poi} as quantisability equations for classical metrics. These are not Einstein's equations but they do involve curvature constraints and may provide the beginning of a mechanism for how the former might yet emerge as a quantum geometry consistency condition. Noncommutativity can also force the quantum calculus to be higher dimensional (as in our case) which is a further constraint studied in \cite{Ma:rec} but not yet analysed at the Poisson level. Returning to our model, the quantum differential structure on $\R\times\Z_n$ (this is independent of any metric) can also be expected to limit the possible diffeomorphisms in the classical limit since at the discrete level it will not be so easy to transform a discrete coordinate to a continuous one, and this may also relate indirectly to the absence of shift vectors in the allowed metric. In fact, diffeomorphism-invariance in GR enters in two ways, as freedom in the choice of coordinates, and as an active symmetry generated by Lie derivatives. The first aspect is taken care of in quantum geometry as the equations and concepts are all coordinate-independent -- we are free to describe our algebras and differential forms with whatever generators and bases we prefer. This means there is a large but classical automorphism group behind any model. In quantum geometry, one has quantum vector fields as right $A$-module maps $\Omega^1\to A$ but these now generate something of a different character. Classically, functions and vector fields together generate the algebra of differential operators on a smooth manifold, which forms a Hopf algebroid as an infinitesimally-generated version of the path groupoid\cite{Xu}. The quantum version is studied in \cite[Chap. 6]{BegMa:book} and specifically in a quantum Hopf algebroid version in \cite{Gho}, which includes results for the finite group case. Another starting point is the notion of a universal measuring bialgebra of any algebra $A$\cite{Ma:dif}, which is much bigger than the classical automorphism group and which has in principle a differential algebra version. Key  to the latter is the concept of a differentiable (co)action, with some related first results in \cite{AziMa}. The physical application of these concepts to quantum geometry and their role in quantum gravity, however, remains very much to be explored.

The rest of the paper focussed on the special case of the $\R\times\Z_n$ FLRW-type cosmological model as a background quantum geometry, where the metric on $\Z_n$ is constant but with an overall variable $\cp(t)$ factor. The Friedmann equations for $\cp(t)$ turned out to be the same as those for the standard flat 1+2 dimensional FLRW model, which is perhaps not too surprising given that the calculus on $\Z_n$ is 2-dimensional. For a natural model with spatial curvature, one could next take a non-Abelian group such as $S_3$ or a fuzzy sphere $\C_\lambda[S^2]$ as in \cite{LirMa} for the spatial sector, to be  considered elsewhere. In the spirit of Connes' approach to internal symmetries of particle physics by tensoring a classical spacetime by a finite-dimensional algebra such as matrices or quaternions\cite{CCM}, one could also consider one of these in place of $\Z_n$, but now from an FLRW perspective.  Note also that for the equations of state for the FLRW model on $\R\times\Z_n$ in Section~\ref{secFRW}, we considered only the  standard form of stress energy tensor for an incompressible fluid. Stress tensors in quantum geometry remain poorly understood, with no general theory. In particular, one can check that the obvious choice
 \begin{equation} T=\extd \phi\tens\extd \phi -  {1\over 2}((\extd\phi,\extd\phi)+ m^2 \phi^2) g\end{equation} 
is not conserved for a free scalar field obeying the Klein-Gordon equation for the geometric Laplacian (\ref{LapFL}). Therefore, it remains to consider further what would be natural as stress tensor for a scalar field, even in our FLRW-type background. Similarly, the Einstein tensor remains poorly understood and while the usual formula ${\rm Ricci}-{1\over 2}g S$ in terms of the Ricci tensor and scalar  was sufficient for our model in Section~\ref{secFRW}, and has also been used in other approaches such as \cite{ADMW}, this is not derived as part of a noncommutative geometric calculus of variations and hence not directly connected to an Einstein-Hilbert action built from the Ricci scalar. Such a calculus of variations appears to be a hard problem and our approach here is to continue to explore both sides so as to first gain experience from specific models, i.e. quantum gravity using a functional integral approach and ideas for the Einstein tensor on a quantum geometry background. The continuum limit even of the $\Z_n$ model, once better understood as discussed above and with general metrics, will provide further input here. A long term goal is to have a parallel Hamiltonian quantisation formalism for quantum gravity, and again this would be interesting even for our baby $\Z_n$ model. 

We then analysed quantum field theory and particle creation in the $\R\times \Z_n$ FLRW case, taking as model the set up of Parker\cite{Par69,Par12,ParNav,ParTom,Par65} applied to  $\R\times S^1$. The main difference compared to the circle case is that we found adiabatic no particle creation solutions  for $\cp(t)$  at $m=0$, not only at $m=\infty$. Another difference of course is that the particle creation $\<N_k\>$ from constant `in' to constant `out' metrics is periodic in the discrete momentum  $k$ rather than decaying as $k\to \infty$ as it would on $S^1$, see Figure~\ref{fig:pol-circle}. This is not surprising since the discrete momentum on $\Z_n$ differs fundamentally in being periodic mod $n$. In principle, one could consider particle creation between the new $m=0$ solutions, but this would need new ideas beyond the ones used (we would not be able to just adapt the circle case). The fact that the comparable here is particle creation on $\R\times S^1$ and not on something 1+2-dimensional reflects that scalar quantum fields themselves are not directly sensitive to the 2D nature of the calculus on $\Z_n$, a situation that we might expect to change for particle creation of higher spin fields. 

Finally, while we have focussed on the quantum field theory, one could consider the quantum mechanics limit. In the flat warm-up case of Section~\ref{secqftR} and following the usual steps of factoring into a wave in the time direction and a slowly varying factor, and adding a potential $V(t,i)$, gives the Schr\"odinger-like equation
\begin{equation}
	\imath \del_t \psi(t,i) = -\frac{1}{\cp^2 m}(\del_+ + \del_-)\psi(t,i) + V(t,i)\psi(t,i). 
\end{equation}
The free particle plane-waves are clearly $\psi_k(t,i)=e^{-\imath E_k t}e^{{2\pi\imath \over n}ik}$ with energy spectrum $E_k = \frac{4}{m\cp^2}\sin^2{(\frac{\pi}{n}k)}$,  for $k=0, \dots, n-1$ so that the trace of the free Hamiltonian is $\sum_{k=0}^{n-1}E_k=  \frac{2n}{m\cp^2}$, compared to the circle case where the trace diverges.  This discrete-space quantum mechanics could be studied further with specific potentials $V(t,i)$. 

In summary, we have indicated several directions for further work building on the results in the present paper. Stepping back, the machinery of quantum Riemannian geometry\cite{BegMa:book} can be applied in principle to almost any unital algebra in a step by step fashion and hence explored in a similar way for other algebras of interest. We have already mentioned \cite{LirMa} as another model, and we refer to the conclusions of \cite{Ma:sq} for further discussion of different algebras that could be interesting.

\appendix

\section{Non $*$-preserving solutions}

We have rightly focussed in the Section~\ref{secpoly} on the unitary or $*$-preserving quantum geometries over $\C$ on $\Z_n$. However, the underlying classification
was done by computer algebra and works over any field of characteristic zero. For completeness, we list the remaining solutions which over $\C$ would not obey the unitarity or `reality' condition (\ref{nablastar}). These could be useful in other contexts over $\R$ or applied to other fields, for example to obtain `digital' quantum geometries over $\F_2$ in the setting of \cite{MaPac} (in this case there could be other solutions also, as the field then has non-zero characteristic). 

{\bf For $n\ge 3$ odd}, there are two further independent solutions:
\begin{align*}
{\rm (i)}\qquad\qquad\qquad	\sigFun{+}{+} &= -\rho \twoForm{+}{+}, \quad
	\sigFun{-}{+} = -\twoForm{+}{-} -2\twoForm{-}{+}, \\
	\sigFun{+}{-} &= \twoForm{-}{+},  \quad
	\sigFun{-}{-} = R_{-}^2(\rho^{-1}) \twoForm{-}{-},	
\end{align*}
giving the geometric structures
\begin{align*}
	\nabla e^+ &= (1+\rho) \twoForm{+}{+},\quad \nabla e^- = (1-R_{-}^{2}(\rho^{-1})) \twoForm{-}{-} + 2(\twoForm{+}{-} + \twoForm{-}{+}) \\
	R_{\nabla}e^+ &= - \partial_{-}(\rho) e^+\wedge \twoForm{-}{+}, \\
	R_{\nabla}e^- &= - \partial_{-}(R_{-}(\rho^{-1})) e^+\wedge \twoForm{-}{-} - 2(1-R_{-}(\rho))e^+\wedge \twoForm{-}{+},\\
	{\rm Ricci} &= \frac{1}{2}\left( -\partial_{-}(R_{-}(\rho)) \twoForm{-}{+} + 2(1-\rho)\twoForm{+}{+} + \partial_{-}(\rho^{-1}) \twoForm{+}{-}  \right), \\
	S &= \frac{1}{2}\left( \frac{\partial_{-}(\rho^{-1})}{a} - \frac{\partial_{-}(R_{-}(\rho))}{R_{-}a} \right), \\
	\Delta f &= \frac{1}{a}(R_{-}{f} - R_{+}({f}))(R_{-}({\rho}) +1).
\end{align*}
{\bf For $n=3$}, we may freely add a map $\alpha$ given by $\alpha(e^-)=\lambda R_+(a)e^+\tens e^+$ to $\nabla e^-$ for a free parameter $\lambda$, and $\alpha(e^+)=0$, so no change to $\nabla e^+$. This agrees with the triangle analysis in \cite[Ex.~8.19]{BegMa:book} aside from a different definition of $\rho$.
\begin{align*}
{\rm (ii)}\qquad\qquad\qquad	\sigFun{+}{+} &= \rho \twoForm{+}{+}, \quad
	\sigFun{+}{-} = -2 \twoForm{+}{-} - \twoForm{-}{+}, \\
	\sigFun{-}{+} &= \twoForm{+}{-}, \quad
	\sigFun{-}{-} = -R_{-}^2(\rho^{-1}) \twoForm{-}{-},
\end{align*}
giving the geometric structures
\begin{align*}
	\nabla e^+ &= (1-\rho)\twoForm{+}{+} + 2(e^+\otimes e^- + \twoForm{-}{+}),\quad \nabla e^- =  (1+R_{-}^{2}(\rho^{-}))\twoForm{-}{-}, \\
	R_{\nabla}e^+ &= -\partial_{-}\rho e^+\wedge \twoForm{-}{+} + 2(1-R_{-}(\rho^{-1}))e^+\wedge\twoForm{-}{-}, \\
	R_{\nabla}e^- &= -\partial_{-}(R_{-}(\rho^{-1}))e^+\wedge\twoForm{-}{-}, \\
	{\rm Ricci} &= \frac{1}{2}\left( -\partial_{-}(R_{-}(\rho)) \twoForm{-}{+} + 2(1-R_{-}^2(\rho^{-1}) )\twoForm{-}{-} + \partial_{-}(\rho^{-1}) \twoForm{+}{-} \right), \\
	S &= \frac{1}{2}\left( \frac{\partial_{-}(\rho^{-1})}{a} - \frac{\partial_{-}(R_{-}(\rho))}{R_{-}a} \right), \\
	\Delta f &= \frac{1}{a}(R_{+}(f) - R_{-}(f))(R_{-}(\rho) +1). 
\end{align*}
{\bf For $n=3$}, we may freely add a map $\alpha$ given by $\alpha(e^+)=\lambda R_+(a)e^-\tens e^-$ to $\nabla e^+$ for a free parameter $\lambda$, and $\alpha(e^-)=0$, so no change to $\nabla e^-$. This again agrees with the triangle analysis in \cite{BegMa:book} aside from a different definition of $\rho$. 

{\bf For $n\ge 4$ even}, there are two further independent solutions each with a free nonzero parameter $q$, from which we define a function
\begin{align*}
	Q = q^{(-1)^{i}} = \begin{pmatrix} q \\ q^{-1} \\ \vdots \end{pmatrix}. 
\end{align*}
Then 
\begin{align*}
{\rm (i)}\qquad\qquad\qquad	\sigma(\twoForm{+}{+}) &= \rho \twoForm{+}{+}, \quad
	\sigma(\twoForm{+}{-}) = (Q-1) \twoForm{+}{-} + Q\twoForm{-}{+},\\
	\sigma(\twoForm{-}{+}) &= \twoForm{+}{-}, \quad
	\sigma(\twoForm{-}{-}) = R_{-}^2(\rho^{-1})Q \twoForm{-}{-},	
\end{align*}
giving the geometric structures
\begin{align*}
	\nabla e^+ &= (1-\rho)\twoForm{+}{+} + (1-Q)(\twoForm{-}{+} + \twoForm{+}{-}),\quad \nabla e^- = (1-R_{-}^2(\rho^{-1})Q) \twoForm{-}{-}, \\
	R_{\nabla}e^+ &= \partial_{-}(\rho R_{+}(Q)) e^+\wedge\twoForm{-}{+} + (R_{+}(Q-1)R_{-}(\rho^{-1}) - (Q-1))e^+\wedge\twoForm{-}{-}, \\
	R_{\nabla}e^- &= \partial_{-}( R_{-}(\rho^{-1})R_{+}(Q)) e^+\wedge \twoForm{-}{+}, \\
	{\rm Ricci} &= \frac{1}{2}\left(\partial_{-}(R_{-}(\rho)Q) \twoForm{-}{+} + \partial_{+}(R_{+}(Q)R_{-}(\rho^{-1})) \twoForm{+}{-}  +((Q-1)R_{-}^2(\rho^{-1}) -R_{-}(Q-1))\twoForm{-}{-} \right), \\
	S &= \frac{1}{2a}\left(\partial_{+}(R_{+}(Q)R_{-}(\rho^{-1})) - R_-(\rho)\partial_{-}(R_{-}(\rho)Q) \right),\\
	\Delta f&=  -\left( \frac{1}{R_{-}(a)} + \frac{1}{a} \right) ( \partial_{-}f + Q\partial_{+}f).
\end{align*}

\begin{align*}
{\rm (ii)}\qquad\qquad\qquad	\sigma(\twoForm{+}{+}) &= \rho Q \twoForm{+}{+}, \quad
	\sigma(\twoForm{-}{-}) = R_{-}^2(\rho^{-1}) \twoForm{-}{-}, \\
	\sigma(\twoForm{+}{-}) &= \twoForm{-}{+}, \quad
	\sigma(\twoForm{-}{+}) = Q\twoForm{+}{-} +(Q-1)\twoForm{-}{+},
\end{align*}
giving the geometric structures
\begin{align*}
	\nabla e^+ &= (1-\rho Q) \twoForm{+}{+},\quad \nabla e^- = (1-R_{-}^2(\rho^{-1})) \twoForm{-}{-} + (1-Q)( \twoForm{+}{-} + \twoForm{-}{+}), \\
	R_{\nabla}e^+ &= \partial_{-}(\rho Q) e^+\wedge\twoForm{-}{-}, \\
	R_{\nabla}e^- &= (-R_{+}(Q-1)R_{-}(\rho) + Q-1)e^+\wedge\twoForm{-}{+} +  \partial_{-}(Q R_{-}(\rho^{-1})) e^+\wedge\twoForm{-}{-}, \\
	{\rm Ricci} &= \frac{1}{2}\left(\partial_{-}(R_{-}(\rho Q)) \twoForm{-}{-} - (\partial_{-}(R_{+}(Q)\rho^{-1}) \twoForm{+}{-} + (\rho(Q-1) - R_{+}(Q-1))\twoForm{+}{+} \right), \\
	S &= -\frac{1}{2 a} \partial_{-}(R_{+}(Q) \rho^{-1}), \\
	\Delta f&=  -\left( \frac{1}{R_{-}(a)} + \frac{1}{a} \right) ( Q\partial_{-}f + \partial_{+}f). 
\end{align*}

{\bf For $n=4$}, we have a further more general form for the generalised braiding
\begin{align*}
	\sigFun{+}{+} = \sigma_0 \twoForm{+}{+} + \sigma_6 \twoForm{-}{-} , \quad \sigFun{+}{-} = \sigma_1 \twoForm{+}{-} + \sigma_2 \twoForm{-}{+}, \\
	\sigFun{-}{+} = \sigma_3 \twoForm{+}{-} + \sigma_4 \twoForm{-}{+}, \quad	\sigFun{-}{-} = \sigma_5 \twoForm{-}{-}+ \sigma_7 \twoForm{+}{+} 
\end{align*}
for which the conditions for zero torsion are the same as before but metric compatibility now has a more complicated form due to the two extra parameters $\sigma_6,\sigma_7$. The QLCs turn out to fall into 10 families of which 3 are the ones with  $\sigma_6 = \sigma_7 = 0$ already covered above.  {\em In addition} we have

(i) a 4-parameter solution  with a free nonzero function $\gamma=(\gamma_0, \gamma_1, \gamma_2, \gamma_3)$ and 
\begin{align*}
	\sigFun{+}{+} &=  \gamma \twoForm{-}{-} , 
	\quad \sigFun{+}{-} =  -\twoForm{+}{-},  \\
	\sigFun{-}{+}& =  -\twoForm{-}{+}, 
	\quad	\sigFun{-}{-} =  R_-(\gamma^{-1})R_+(\rho')  \twoForm{+}{+}, \\
	\nabla e^+ &=  \twoForm{+}{+} + \twoForm{-}{+} + \twoForm{+}{-} - \gamma\twoForm{-}{-},\\
	\nabla e^- &= \twoForm{-}{-} + \twoForm{+}{-} + \twoForm{-}{+} - R_-(\gamma^{-1})R_{+}(\rho')\twoForm{+}{+},
\end{align*}
where
\[ \rho'={1\over \rho R_{+}\rho}.\]
This is $*$-preserving if and only if $\gamma$ has the 2-parameter form  such that $R_+^2(\gamma)=\bar\gamma^{-1}$ as in the main text. 

(ii) a 3-parameter solution  with parameter $\beta$ and functions 
\[\gamma=(p, q, p, q), \quad \delta={pq-1\over R_+(\gamma)-1}=(pq-1)({1\over q-1}, {1\over p-1},{1\over q-1},{1\over p-1}),\]
\begin{align*}
	\sigFun{+}{+} =  \rho (1-\delta) \twoForm{+}{+} + \beta (\gamma-1) \rho' \twoForm{-}{-}, \quad \sigFun{+}{-} =  (\gamma-1)\twoForm{+}{-} + \gamma \twoForm{-}{+}, \\
	\sigFun{-}{+} =  (1-\delta) \twoForm{+}{-} - \delta \twoForm{-}{+}, \quad 
	\sigFun{-}{-} =  -{  \delta\over \beta R_+^2\rho' }\twoForm{+}{+}  + {\gamma\over R^{2}_{+}\rho}\twoForm{-}{-},
\end{align*}
where
\[ 
\rho' = (\frac{\rho_0}{\rho_2},\rho_0\rho_{1},1,\rho_0\rho_3),
\]
giving the QLC 
\begin{align*}
	\nabla e^+ &= (1 - \rho (1-\delta)) \twoForm{+}{+} + (1-\gamma)(\twoForm{-}{+} + \twoForm{+}{-}) + \beta \rho'(1-\gamma) \twoForm{-}{-}, \\
	\nabla e^- &= (1-{\gamma\over R^{2}_{+}\rho})\twoForm{-}{-} + \delta(\twoForm{+}{-} +\twoForm{-}{+}) + {\delta\over\beta R_+^2\rho'}\twoForm{+}{+}.\\
\end{align*}

(iii) a 3-parameter solution  with parameters $ \beta$ and  functions 
\[\gamma = ( p, 0,  q, 0), \quad \delta = (1,{q\over p},1,{p\over q}),\],
\begin{align*}
	\sigFun{+}{+} &=    R_-\left({\gamma\over \gamma-1}\right) \rho\twoForm{+}{+}+ { \beta \delta \rho' \over 1-R_-(\gamma) } \twoForm{-}{-} , \\ 
	\sigFun{+}{-} &=  ( \gamma-1)\twoForm{+}{-} +  \gamma \twoForm{-}{+}, \\
	\sigFun{-}{+} &=  R_{+}\left({\gamma \over \gamma-1}\right)\twoForm{+}{-} +  \frac{1}{R_{+}(\gamma-1)}\twoForm{-}{+}, \\
	\sigFun{-}{-} &=  { R_-(\delta) \over \beta R_+^2(\rho')}(1-\gamma)\twoForm{+}{+} +  R^{2}_{+}({\gamma\over\rho})\twoForm{-}{-},
\end{align*}
where
\[ \rho'= ({\rho_0\over\rho_2}, \rho_0\rho_1 , 1 , \rho_0\rho_3), \]
giving the QLC 
\begin{align*}
	\nabla e^+ &=  (1+R_-({ \gamma \over 1-\gamma})\rho)  \twoForm{+}{+} + (1-\gamma)(\twoForm{-}{+} + \twoForm{+}{-} ) -   {\beta \delta \rho' \over 1 - R_-(\gamma)} \twoForm{-}{-},  \\
	\nabla e^- &= \left(1-R^{2}_{+}\left({\gamma\over\rho}\right)\right)\twoForm{-}{-} + {1 \over 1-R_{+}(\gamma)}(\twoForm{+}{-}+\twoForm{-}{+}) - { R_-(\delta) \over \beta R_+^2\rho'}(1-\gamma)\twoForm{+}{+}.
\end{align*}

(iv) a 3-parameter solution  with parameters $\beta$ and the functions 
\[\gamma = ( 0,  p, 0,  q ), \quad \delta = ({p\over q},1,{q\over p},1),\], 
\begin{align*}
	\sigFun{+}{+} &=  \rho  R_-({\gamma\over \gamma-1}) \twoForm{+}{+}+ { \beta\delta \rho' \over 1-R_-(\gamma) } \twoForm{-}{-} , \\ 
	\sigFun{+}{-} &=  ( \gamma-1)\twoForm{+}{-} +  \gamma \twoForm{-}{+}, \\
	\sigFun{-}{+} &=  R_{+}(\frac{\gamma}{\gamma-1})\twoForm{+}{-} +  \frac{1}{R_{+}(\gamma-1)}\twoForm{-}{+}, \\
	\sigFun{-}{-} &=  { R_-(\delta) \over   \beta R_+^2(\rho')}(1-\gamma)\twoForm{+}{+} +  R^{2}_{+}({\gamma\over\rho})\twoForm{-}{-}, 
\end{align*}
where
\[    \rho'= ({\rho_0\over\rho_2} , \rho_0\rho_1  , 1, \rho_0\rho_3 ),  \]
giving the QLC 
\begin{align*}
	\nabla e^+ &=  (1+R_-({\gamma \over 1-\gamma})\rho)  \twoForm{+}{+} + (1-\gamma)(\twoForm{-}{+} + \twoForm{+}{-} ) -   {\beta \delta\rho' \over 1 - R_-(\gamma)} \twoForm{-}{-},  \\
	\nabla e^- &= (1-R^{2}_{+}({\gamma\over\rho}))\twoForm{-}{-} + {1\over1-R_{+}(\gamma)}(\twoForm{-}{+}+\twoForm{-}{+}) - { R_-(\delta) \over \beta R_+^2\rho'}(1-\gamma)\twoForm{+}{+}. 
\end{align*}

(v) a 2-parameter solution  with parameter $\beta$ and $Q= ( q, q^{-1}, q, q^{-1} )$ as usual, 
\begin{align*}
	\sigFun{+}{+} =  \rho \twoForm{+}{+}, \quad 
	\sigFun{+}{-} =  (Q-1)\twoForm{+}{-} +  Q\twoForm{-}{+},\\
	\sigFun{-}{+} =  \twoForm{+}{-}, \quad
	\sigFun{-}{-} =  \beta  \rho' \twoForm{+}{+} +  R^{2}_{+}(\rho^{-1})Q\twoForm{-}{-},
\end{align*}
where
\[  \rho'= ( 1, -\frac{\rho_1\rho_2}{q} , {\rho_2\over\rho_0} , -\frac{\rho_2\rho_3}{q}  ), \]
giving the QLC 
\begin{align*}
	\nabla e^+ &=  (1-\rho)\twoForm{+}{+} + (1-Q)(\twoForm{+}{-} + \twoForm{-}{+}),   \\
	\nabla e^- &=   (1-R^{2}_{+}(\rho^{-1})Q)\twoForm{-}{-} -\beta \rho' \twoForm{+}{+}. 
\end{align*}

(vi) a 2-parameter solution  with parameter $\beta$ and $Q= ( q, q^{-1}, q, q^{-1} )$ as usual,
\begin{align*}
	\sigFun{+}{-}& =   \twoForm{-}{+},\quad \sigFun{-}{+} =  Q\twoForm{+}{-} +  (Q-1)\twoForm{-}{+},\\
	\sigFun{+}{+} &=  \rho Q\twoForm{+}{+},\quad \sigFun{-}{-} =  \beta \rho'\twoForm{+}{+} +  R^{2}_{+}(\rho^{-1})\twoForm{-}{-}, 
\end{align*}
where
\[  \rho'= ( 1, -\frac{\rho_1\rho_2}{q} , {\rho_2\over\rho_0} , -\frac{\rho_2\rho_3}{q}  ), \]
giving the QLC 
\begin{align*}
	\nabla e^+ &= (1-\rho Q)\twoForm{+}{+}, \\
	\nabla e^- &= (1-R^{2}_{+}(\rho^{-1}))\twoForm{-}{-} + (1-Q)(\twoForm{+}{-} +\twoForm{-}{+}) - \beta \rho' \twoForm{+}{+}. 
\end{align*}

(vii) a 2-parameter solution  with parameter $\beta$ and $Q= ( q, q^{-1}, q, q^{-1} )$ as usual,
\begin{align*}
	\sigFun{+}{+} = - \rho'\rho Q \twoForm{+}{+}+  \beta\rho''\twoForm{-}{-} , \quad 
	\sigFun{+}{-} =   \twoForm{-}{+}, \\
	\sigFun{-}{+} =  - \rho'Q\twoForm{+}{-} - ( \rho'Q+1)\twoForm{-}{+}, \quad
	\sigFun{-}{-} =   R^{2}_{+}(\rho^{-1})\twoForm{-}{-},
\end{align*}
where
\[   \rho' = (\rho_1\rho_0, \rho^{-1}_0\rho^{-1}_1, \rho_1\rho_0, \rho^{-1}_0\rho^{-1}_1),\quad \rho'' = ({\rho_0\over \rho_2}q,1,q, {\rho_3\over \rho_1}), \]
giving the QLC
\begin{align*}
	\nabla e^+ &=  (1+ \rho'\rho Q)\twoForm{+}{+}  - \beta\rho''\twoForm{-}{-}, \\
	\nabla e^- &= (1 - R^{2}_{+}(\rho^{-1}))\twoForm{-}{-} + (1+ \rho'Q)(\twoForm{+}{-} + \twoForm{-}{+}). 
\end{align*}

Note that $\Z_4$ here is a different group from $\Z_2\times\Z_2$ treated in  \cite{Ma:sq}\cite[Ex.~8.20]{BegMa:book}, even though in both cases the graph is a square. This means that, although $\Omega^1$ and the metric can be made to match up and hence the metric compatibility part of the QLC condition is the same,  $\Omega^2$ and hence the condition for torsion freeness are different. This work \cite{Ma:sq} also treats the $\Z_2$ case.


\begin{thebibliography}{99}
\bibitem{Ma:haw}S. Majid, Quantum Riemannian geometry and particle creation on the integer line, Class. Quantum Grav. 36 (2019) 135011 (22pp) 
\bibitem{Ma:sq}S. Majid, Quantum gravity on a square graph,   Class. Quantum Grav 36 (2019) 245009 (23pp)
\bibitem{BegMa:book} E.J. Beggs and S. Majid, {\em Quantum Riemannian Geometry}, Grundlehren der mathematischen Wissenschaften, Vol. 355, Springer (2020) 809pp.
\bibitem{Ma:gra} S. Majid, Noncommutative Riemannian geometry of graphs, J. Geom. Phys. 69 (2013) 74--93
\bibitem{BegMa:gra}
 E.J. Beggs and S. Majid, Gravity induced by quantum spacetime, Class. Quantum. Grav. 31 (2014) 035020 (39pp)
 \bibitem{MaTao}S. Majid and W.-Q. Tao, Cosmological constant from quantum spacetime, Phys. Rev. D  91 (2015)  124028 (12pp)
\bibitem{MaPac}S. Majid and A. Pachol, Digital finite quantum Riemannian geometries, J. Phys. A 53 (2020) 115202 (40pp) 
\bibitem{Loll}J. Ambjorn, J. Jurkiewicz  and R. Loll, Dynamically triangulating Lorentzian quantum gravity, Nucl. Phys. B610 (2001) 347--382
\bibitem{Ash}A. Ashtekar, T. Pawlowski, P. Singh and K. Vandersloot, Loop quantum cosmology of k=1 FRW Models, Phys. Rev. D 75 (2007) 024035-1-26
\bibitem{Dow} F. Dowker, Introduction to causal sets and their phenomenology, Gen. Rel. Grav, 45 (2013)1651--1667
\bibitem{Hale}M. Hale, Path integral quantisation of finite noncommutative geometries, J. Geom. Phys. 44 (2002) 115--128
\bibitem{HawGib}S. Hawking and G. Gibbon, eds. {\em Euclidean Quantum Gravity}, World Scientific (1993)
\bibitem{Ma:alm}S. Majid, Almost commutative Riemannian geometry: wave operators, Commun. Math. Phys. 310 (2012) 569--609
\bibitem{Ma:rec}S. Majid, Reconstruction and quantization of Riemannian structures, J. Math. Phys.  61 (2020)  022501 (32pp)
\bibitem{Par69}
L.~Parker, Quantized fields and particle creation in expanding universes. 1, Phys.\ Rev.\  {\bf 183} (1969) 1057
\bibitem{Par12} L.~Parker,  Particle creation and particle number in an expanding universe, J.\ Phys.\ A {\bf 45} (2012) 374023 
\bibitem{ParNav}L. Parker and J. Navarro-Salas, Fifty years of cosmological particle creation, arXiv:1702.07132 (physics.hist-ph)
\bibitem{ParTom} L. Parker and D. Toms, {\em Quantum Field Theory in Curved Spacetime: Quantized Fields and Gravity}, Cambridge University Press (2009)
\bibitem{MukWin}V. Mukhanov and S. Winitzki, {\em Introduction to Quantum Effects in Gravity}, Cambridge University Press (2007)
\bibitem{Birrel} N.D. Birrell and P.C.W. Davies, {\em Quantum Fields in Curved Space}, Cambridge University Press (1984)
\bibitem{Ma:pla}S. Majid, Hopf algebras for physics at the Planck scale, Class. Quantum Grav. 5 (1988) 1587--1607
\bibitem{MaRue}S. Majid and H. Ruegg, Bicrossproduct structure of the $\kappa$-Poincar\'e group and non-commutative geometry, Phys. Lett. B. 334 (1994) 348--354
\bibitem{Luk}J. Lukierski, H. Ruegg, A. Nowicki and V.N. Tolstoi,  q-Deformation of Poincare algebra,  Phys. Lett. B 264 (1991) 331
\bibitem{Sny}H.S. Snyder, Quantized space-time, Phys. Rev. D 67 (1947) 38--41
\bibitem{DFR} S. Doplicher, K. Fredenhagen and J. E. Roberts, The quantum structure of spacetime at the Planck scale and quantum fields, Commun. Math. Phys. 172 (1995) 187--220
\bibitem{Hoo}G. 't Hooft, Quantization of point particles in 2+1 dimensional gravity and space- time discreteness, Class. Quant. Grav. 13 (1996) 1023
\bibitem{Con} A. Connes, {\em Noncommutative Geometry}, Academic Press (1994).
\bibitem{DVM}M. Dubois-Violette and P.W. Michor, Connections on central bimodules in noncommutative differential geometry, J. Geom. Phys. 20 (1996) 218--232
\bibitem{Mou}J. Mourad, Linear connections in noncommutative geometry, Class. Quantum Grav. 12 (1995) 965--974
\bibitem{sagemath}Sage Developers, {\em The Sage Mathematics Software System (Version 8.4)}, 2020, https://www.sagemath.org.
 \bibitem{LirMa} E. Lira Torres and S. Majid, {\em Quantum gravity and Riemannian geometry on the fuzzy sphere}, arXiv:2004.14363 (math.QA) 
 \bibitem{Par65} L.~Parker, On the magnetic moment of a charged particle in a changing magnetic field, Nuovo Cimento 40B (1965) 99
\bibitem{BegMa:geo}E.J. Beggs and S. Majid, Quantum geodesics in quantum mechanics, arXiv:1912.13376 (math-ph)
 \bibitem{BegMa:poi} E.J. Beggs and S. Majid, Poisson-Riemannian geometry, J. Geom. Phys.  114 (2017) 450--491 
\bibitem{Xu}P. Xu, Quantum groupoids, Commun. Math. Phys., 216 (2001) 539--581
\bibitem{Gho}A. Ghobadi, Hopf algebroids, bimodule connections and noncommutative geometry, arXiv:2001.08673
\bibitem{Ma:dif}S. Majid, Quantum and braided diffeomorphism groups, J. Geom. Phys. 28 (1998) 94--128
\bibitem{AziMa}R. Aziz and S. Majid, Quantum differentials on cross product Hopf algebras, J. Algebra 531 (2020) 303--351 
\bibitem{CCM}A. Chamseddine, A. Connes and M. Marcolli, Gravity and the standard model with neutrino
mixing, Adv. Theor. Math. Phys. 11 (2007) 991
\bibitem{ADMW}P. Aschieri, M. Dimitrijevi\'c, F. Meyer and J. Wess, Noncommutative geometry and gravity, Class. Quant. Grav. 23 (2006) 1883





\end{thebibliography}
\end{document}